\documentclass[usenatbib]{mnras}
\usepackage{graphicx}
\usepackage{amssymb, amsmath}
\usepackage{times}
\usepackage{color}
\usepackage{multirow}
\usepackage{lscape}
\usepackage{varwidth}
\usepackage[section]{placeins}
\usepackage{epstopdf}
\usepackage{epsf}
\usepackage{float}
\usepackage{booktabs}
\usepackage[normalem]{ulem}

\newcommand{\cmnt}[1]{}

\newcommand{\HI}{\ion HI}
\newcommand{\lya}{Lyman-$\alpha$ }

\begin{document}
\setlength{\parskip}{0pt}

\title[]{Impact of inhomogeneous reionization on post-reionization 21 cm intensity mapping measurement of cosmological parameters}

\author[Long et. al]{Heyang Long,$^{1,2}$\thanks{E-mail: long.1697@osu.edu} Catalina Morales-Gutiérrez,$^{3,4}$ \thanks{Email: catalina.moralesgutierrez@ucr.ac.cr} Paulo Montero-Camacho,$^{5,6}$\thanks{E-mail: pmontero@pcl.ac.cn} and\newauthor
Christopher M. Hirata$^{1,2,7}$\thanks{E-mail: hirata.10@osu.edu}\\
$^1$Department of Physics, The Ohio State University, 191 West Woodruff Avenue, Columbus, Ohio 43210, USA\\
$^2$Center for Cosmology and AstroParticle Physics (CCAPP), The Ohio State University, 191 West Woodruff Avenue, Columbus, Ohio 43210, USA\\
$^3$Department of Physics, University of Costa Rica, 11501 San José, Costa Rica.\\
$^4$Space Research Center (CINESPA), University of Costa Rica, 11501 San José, Costa Rica.\\
$^5$Department of Mathematics and Theory, Peng Cheng Laboratory, Shenzhen, Guangdong 518066, China\\
$^6$Department of Astronomy, Tsinghua University, Beijing 100084, China \\
$^7$Department of Astronomy, The Ohio State University, 140 West 18th Avenue, Columbus, Ohio 43210, USA}

\date{\today}

\maketitle

\begin{abstract}
21 cm intensity mapping (IM) has the potential to be a strong and unique probe of cosmology from redshift of order unity to redshift potentially as high as 30. For post-reionization 21 cm observations, the signal is modulated by the thermal and dynamical reaction of gas in the galaxies to the passage of ionization fronts during the Epoch of Reionization. In this work, we investigate the impact of inhomogeneous reionization on the post-reionization 21 cm power spectrum and the induced shifts of cosmological parameters at redshifts $3.5 \lesssim z \lesssim 5.5$. We make use of hydrodynamics simulations that could resolve small-scale baryonic structure evolution to quantify \HI\ abundance fluctuation, while semi-numerical large box {\sc 21cmFAST} simulations capable of displaying inhomogeneous reionization process are deployed to track the inhomogeneous evolution of reionization bubbles. We discussed the prospects of capturing this effect in two post-reionization 21 cm intensity mapping experiments: SKA1-LOW and PUMA. We find the inhomogeneous reionization effect could impact the \HI\ power spectrum up to tens of percent level and shift cosmological parameters estimation from sub-percent to tens percent in the observation of future post-reionization 21 cm intensity mapping experiments such as PUMA, while SKA1-LOW is likely to miss this effect at the redshifts of interest given the considered configuration. In particular, the shift is up to 0.0206 in the spectral index $n_s$ and 0.0192 eV in the sum of the neutrino masses $\sum m_\nu$ depending on the reionization model and the observational parameters. We discuss strategies to mitigate and separate these biases.
\end{abstract}


\section{Introduction}

Intensity mapping of the neutral hydrogen (\HI) 21 cm hyperfine line has long been recognized as a potentially unparalleled probe of cosmology \citep{2006PhR...433..181F}.  At low redshifts, after the epoch of reionization (EoR), the intensity mapping (IM) signal is expected to come mainly from gas in galaxies, whose unresolved emission traces large scale structure. In this regime, 21 cm IM can be used as a large-scale structure tracer, including for measuring the cosmic expansion history with baryon-acoustic oscillations (BAOs; \citealt{2008MNRAS.383.1195W, 2008PhRvL.100i1303C}) and measuring the growth of structure (although due to the mean signal degeneracy, this requires more techniques than just the redshift space quadrupole; \citealt{2018JCAP...05..004O, 2019JCAP...06..025C, 2019JCAP...07..023C}). While at the lowest redshifts, the 21 cm could provide an independent measurement from optical spectroscopic surveys, or be used for cross-correlations \citep{2009MNRAS.394L...6P, 2010Natur.466..463C, 2013MNRAS.434L..46S, 2022arXiv220201242C}, IM becomes a particularly promising tool at $2<z<6$, where there is an enormous volume with a large number of individually faint galaxies. (The comoving volume at $2<z<6$ is tripled compared to that at $0<z<2$.)
Extending exploration to this redshift range could characterize the expansion history of the Universe and growth rate of structure to the matter-dominated, pre-acceleration era. The detection could provide new tests to the minimal $\Lambda$CDM model as well as put constraints on alternatives such as early dark energy \citep{2006JCAP...06..026D, 2016PhRvD..94j3523K}. One might also use the large number of modes to search for inflationary relics in the primordial power spectrum or bispectrum \citep{2013PhRvL.111q1302C, 2014JCAP...03..039M,2018arXiv181009572C}. At higher redshifts, one may probe the epoch of reionization ($6\lesssim z\lesssim 12$), when early star-forming galaxies and quasars emitted photons with energy high enough to ionize the hydrogen throughout the Universe that had been neutral since the time of recombination in the intergalactic medium (IGM). Beyond this, one may trace back to ``Cosmic Dawn'', in order to learn about the first sources of ionization and heating in the mostly-neutral Universe \citep{2015aska.confE...1K, 2016PhR...645....1B}. The highest redshifts probe undisturbed primordial gas, and likely again contain enormous cosmological information due to the enormous volume and tiny Jeans scale of this gas, but with the disadvantage of extraordinarily bright foregrounds \citep{2004PhRvL..92u1301L, 2004NewA....9..417P}. In this work, we will focus on cosmological measurements with post-EoR 21 cm IM experiments, i.e., between $2<z<6$.

Several ongoing radio telescope projects that conduct \HI\ 21 cm intensity mapping surveys to measure the relatively low-redshift Universe with fainter foregrounds include the Canadian Hydrogen Intensity Mapping Experiment (CHIME; \citealt{2014SPIE.9145E..22B}), Tianlai \citep{2012IJMPS..12..256C}, and the BAO from Integrated Neutral Gas Observations (BINGO; \citealt{2019JPhCS1269a2002W, 2022A&A...664A..14A}), the Hydrogen Intensity and Real-time Analysis eXperiments (HIRAX; \citealt{2016SPIE.9906E..5XN}), the Ooty Wide Field Array (OWFA; \citealt{2017JApA...38...10S}). To extend the detection of 21 cm to post-reionization high-redshift Universe ($2<z<6$), more dedicated experiments have been proposed, e.g., the Square Kilometre Array (SKA; \citealt{2020PASA...37....7S, 2019arXiv191212699B}) and the``Stage II'' Packed Ultrawide-band Mapping Array (PUMA; \citealt{2018arXiv181009572C}) reference concept that could map the Universe up to $z\approx6$. The epoch of reionization itself will be probed with SKA or with dedicated arrays such as the Hydrogen Epoch of Reionization Array (HERA; \citealt{2017PASP..129d5001D}).

To achieve the cosmology goals for experiments such as PUMA in the post-reionization context, it is imperative to thoroughly understand how the physics of the EoR affects the 21 cm signal from galaxies at a lower redshift.
The imprints left by the EoR can be understood in two steps -- the imprint of EoR on the thermal history of the IGM \citep{2019MNRAS.490.3177W,2019MNRAS.486.4075O,2019MNRAS.487.1047M}, and the impact of the IGM on gas accretion into galaxies. During the EoR, IGM gas was photo-heated by the passing ionization fronts to a typical temperature of order $2\times 10^4$ K \citep{1994MNRAS.266..343M, 2008ApJ...689L..81T,2009ApJ...701...94F,2015ApJ...813L..38D,2019ApJ...874..154D}. Afterwards, the gas cooling process was driven primarily by the adiabatic expansion of the Universe, heating from residual recombinations followed by photoionization from the UV background, and (for early enough reionization) inverse Compton cooling \citep{2016MNRAS.456...47M}, which took cosmological timescale for gas to relax. Baryonic structures in some very small scale halos ($\leq 10^7M_{\odot}$) were destroyed during the EoR \citep{2004MNRAS.348..753S, 2018MNRAS.474.2173H}. Afterwards, the growth of small-scale structure in the gas is suppressed, and after hundreds of millions of years this filtering reaches $\sim 100$ kpc scales \citep{2020ApJ...898..149D}. The cut-off halo mass scale below which the baryon fraction is reduced to approximately the universal value can be characterized by the filtering mass \citep{1998MNRAS.296...44G,2000ApJ...542..535G}. The evolution of filtering masses \citep{2022MNRAS.513..117L} with respect to the time of reionization and time passed after it also reflects the hydrodynamic response of IGM to reionization.

Driven by discrete ionizing sources with distinct properties, the reionization process of the Universe is expected to be inhomogeneous. At the begining of the EoR, star-forming galaxies and quasars started emitting photons that could escape from the local ISM and ionize neutral hydrogens in the IGM such that the regions called reionization bubbles were created. Then EoR proceeded by the expansion of these reionization bubbles until they overlapped and saturated the whole Universe. Thus, there is a spread of local reionization times of different regions in the Universe, as termed by \textit{inhomogeneous} or \textit{patchy} reionization. Since the thermal and dynamical relaxation time for IGM could be comparable to the duration of EoR, the post-reionization states of IGM sensitive to the local reionization time also inherited the inhomogeneous signature. Post-reionization IGM state fluctuations due to inhomogeneous reionization could provide the information about EoR itself and the nature of ionizing sources \citep{2020MNRAS.499.1640M, 2022arXiv220713098P}. On the other hand, these fluctuations could enter into the analysis of cosmology measurements, such as \lya\ forest surveys \citep{2006ApJ...644...61L} and high redshift galaxy surveys \citep{2006ApJ...640....1B}. Observations of \lya\ troughs below $z=6$ has given evidence for inhomogeneous reionization even at these late times \citep[e.g.][]{2015MNRAS.447.3402B}. Previous studies regarding inhomogeneous reionization as a systematic effect for cosmological measurement have examined the CMB anisotropy \citep{1998PhRvL..81.2004K}, the anisotropic secondary CMB $B$-mode polarization generated by scattering of CMB photons with free electrons during patchy reionization \citep{2021JCAP...01..003R}, the impact on baryonic content of dwarf galaxies and cosmic filaments \citep{2020MNRAS.494.2200K}, temperature fluctuation of IGM \citep{2008ApJ...689L..81T}, the impact on the \lya\ forest power spectrum \citep{2019MNRAS.487.1047M,2022ApJ...928..174M, 2022MNRAS.509.6119M}, and the impact on the galaxy power spectrum \citep{2006ApJ...640....1B}.  There is ongoing work to include patchy reionization in numerical simulations \citep{2022arXiv220304337C, 2022arXiv220713098P} and study the impact.

In this work, we investigate the imprints of inhomogeneous reionization on the cosmological parameter measurements of post-reionization 21 cm IM experiments. In \S\ref{sec:formalism}, we present the formalism to calculate the total 21 cm power spectrum taking into account the impact of inhomogeneous reionization and the Fisher matrix framework to calculate the parameter shifts due to this effect. In \S\ref{sec:sims}, we layout the large-scale {\sc 21cmFAST} and small-scale GADGET-2 simulation suites we employ in this work. In \S\ref{sec:results}, we display main results for this work: contribution from inhomogeneous reionization to post-reionization 21 cm power spectrum compared to fiducial model, cosmological parameter shifts due to the impact of inhomogenous reionization. In \S\ref{sec:disc}, we discuss the implications of our result, the mitigation strategies for this effect, the possible directions for future work and the caveats of this work.

\section{Conventions and formalism}\label{sec:formalism}
\subsection{Cosmology}
The fiducial cosmological parameters throughout this work  are from the Planck 2015 ``TT+TE+EE+lowP+lensing+ext'' \citep{2016A&A...594A..13P}: 
$h$=0.6774, $\Omega_{\rm c}h^2=0.1188$, $\Omega_{\rm b}h^2=0.02230$, $A_{\rm s}=2.142\times 10^{-9}$, $n_{\rm s}=0.9667$, $\alpha_{\rm s}=-0.002$, $\tau_{\rm reio}=0.066$ and $\sum m_{\nu}=0.194$. For \HI\ abundance $\Omega_{\rm HI}$, we cite values measured in \citet{2015MNRAS.452..217C}.

\subsection{Patchy reionization modeling}

To quantify the contribution from inhomogeneous reionization to the post-reionization \HI\ 21 cm power spectrum, we adopt a method analogous to that of \cite{2019MNRAS.487.1047M} in which the patchy reionization effect on \lya\ forest is studied. We introduce a term $\Xi(z_{\rm re},\,z_{\rm obs}\,|\,\overline{z}_{\rm re})$ in addition to vanilla linear \HI\ biasing model to represent the \HI\ overdensity due to inhomogeneous reionization:
\begin{equation}\label{eq:delta_HI}
    \delta_{\rm HI} = (b_{\rm HI}+\mu^2f)\delta_{\rm m}+\Xi(z_{\rm re},\,z_{\rm obs}\,|\,\overline{z}_{\rm re}),
\end{equation}
where $b_{\rm HI}$ is the \HI\ bias coefficient, $\mu = \cos{\theta}$ is the cosine of angle to the line of sight, $f$ is the linear growth rate, $\delta_{\rm m}$ is the matter overdensity, $z_{\rm re}$ is the local reionization redshift, $\overline{z}_{\rm re}$ is the overall mean reionization redshift (in practice we use the mid-point of reionization $z_{\rm mid}$ as $
\overline{z}_{\rm re}$), $z_{\rm obs}$ is the observation redshift. Specifically, $\Xi$ is written as
\begin{equation}\label{eq:xi}
    \Xi(z_{\rm re},z_{\rm obs}\,|\,\overline{z}_{\rm re}) = \ln{\frac{\rho_{\rm HI}({z_{\rm re},\,z_{\rm obs}})}{\rho_{\rm HI}({\overline{z}_{\rm re},\,z_{\rm obs}})}}.
\end{equation}
Note that when $\rho_{\rm HI}({z_{\rm re},\,z_{\rm obs}})/\rho_{\rm HI}({\overline{z}_{\rm re},\,z_{\rm obs}})\ll1$, 
\begin{equation}
    \ln{\frac{\rho_{\rm HI}({z_{\rm re},\,z_{\rm obs}})}{\rho_{\rm HI}({\overline{z}_{\rm re},\,z_{\rm obs}})}} \approx \frac{\rho_{\rm HI}({z_{\rm re},\,z_{\rm obs}})}{\rho_{\rm HI}({\overline{z}_{\rm re},\,z_{\rm obs}})}-1
\end{equation}
so $\Xi$ captures the small \HI\ density perturbations because of patchy reionization compared to the fiducial homogeneous reionization represented by $\rho_{\rm HI}({\overline{z}_{\rm re},\,z_{\rm obs}})$. Another advantage of using the logarithm form overdensity is that when comparing $\Xi$ with different reionization redshifts, no mean reionization redshift is needed as reference since
\begin{equation}
    \Xi(z_{\rm re,1},z_{\rm obs}\,|\,\overline{z}_{\rm re})
    -\Xi(z_{\rm re,2},z_{\rm obs}\,|\,\overline{z}_{\rm re})
    =\ln{\frac{\rho_{\rm HI}({z_{\rm re,1},\,z_{\rm obs}})}{\rho_{\rm HI}({z_{\rm re,2},\,z_{\rm obs}})}}.
\end{equation}

At some time after the EoR, the gas content of the smaller haloes is suppressed due to the dynamic response of the gas to reionization. In this work, we adopt the methodology based on filtering mass \citep{1998MNRAS.296...44G} tabulated in \cite{2022MNRAS.513..117L} to quantify the small-scale structure effect. The \HI\ density is written as
\begin{equation}\label{eq:rho_HI}
    \rho_{\rm HI}(z_{\rm re},\,z_{\rm obs}) =\int dM_{\rm halo}\frac{dn(M_{\rm halo},z)}{dM_{\rm halo}}M_{\rm HI}(M_{\rm halo},z_{\rm obs}, z_{\rm re})
\end{equation}
where $dn/dM$ is the halo mass function \citep{2008ApJ...688..709T}, $M_{\rm halo}$ is the halo mass, $M_{\rm HI}$ is the H\,{\sc{i}} mass within a halo of mass $M_{\rm halo}$ at observational redshift $z_{\rm obs}$ and reionization redshift $z_{\rm re}$.

The $M_{\rm HI}$ - $M_{\rm halo}$ relation is taken to be approximately linear with a cut-off at filtering mass $M_{\rm F}$. The form of the cutoff is
\begin{equation}\label{eq:HI_mass}
    M_{\rm HI}(M_{\rm halo},z_{\rm obs},z_{\rm re})\propto \frac{f_b M_{\rm halo}}{[1+(2^{1/3}-1)M_F(z_{\rm obs}, z_{\rm re})/M_{\rm halo}]^3},
\end{equation}
where $f_b\equiv\Omega_b/\Omega_m\approx0.1573$ is the universal baryon fraction, and $M_{\rm F}$ values are extracted from a modified {\sc Gadget-2} \citep{2001NewA....6...79S, 2005Natur.435..629S} hydrodynamic simulation as described and tabulated in \cite{2022MNRAS.513..117L}. We have indicated this relation as a proportionality since it is used only to determine $\Xi$: in particular, the absolute normalization is not used and is instead controlled by the parameter $\Omega_{\rm HI}(z)$ in each redshift bin (see \citealt{2018ApJ...866..135V} and discussion in \S\ 3.1 of \citealt{2022MNRAS.513..117L}).

The total 21 cm intensity mapping power spectrum we get based on \HI\ overdensity in Eq.~(\ref{eq:delta_HI}) is composed of the fiducial linear 21 cm power spectrum, the cross-power spectrum contributed by the inhomogeneous reionization, and a thermal noise term $P_{\rm N}$. Note that we do not include shot noise here because as pointed out by \cite{2017MNRAS.471.1788C}, it is negligible for $z > 3$ and at the range of scales we use ($k \sim 0.2h$ Mpc$^{-1}$). The power spectrum is
\begin{eqnarray}
    && \!\!\!\!\!\!\!\!\!\!\!\!\!\!\!\!\!\!\!\!\!\!\!\! P_{21}^{\rm tot}(k,\mu,z_{\rm obs}) 
    \nonumber \\
    &=& \overbrace{\overline{T}^2_{\rm b}(z_{\rm obs})\left(b_{\rm HI}(z_{\rm obs})+\mu^2f\right)^2P_m(k, z_{\rm obs}) +P_{\rm N}}^{\text{\normalsize $P_{21}^{\rm fid}$}} \nonumber \\
    && + \underbrace{2\overline{T}^2_{\rm b}(z_{\rm obs})\left(b_{\rm HI}(z_{\rm obs})+\mu^2f)P_{\rm m,\Xi}(k, z_{\rm obs}\right)}_{\text{\normalsize $P_{21}^{\rm patchy}$}},
\label{eq:P_21}
\end{eqnarray}
where $\overline{T}_{\rm b}$ is the mean brightness temperature \citep{2006ApJ...652..849F, 2012RPPh...75h6901P}, given by
\begin{equation}
    \overline{T}_{\rm b}(z) = 27 \ \textup{mK} \left(\frac{1+z}{10}\right)^{1/2}\left(\frac{\Omega_{\rm HI}(z)h^2}{0.023}\right)\left(\frac{0.15}{\Omega_{\rm m}h^2}\right)^{1/2};
\end{equation}
$P_{\rm m}$ is the linear matter power spectrum calculated by CLASS \citep{2011JCAP...07..034B}; $P_{\rm m,\Xi}$ is the cross-power spectrum between matter overdensity and $\Xi$; and $P_{\rm N}$ is the thermal noise. Our method of developing $P_{\rm m,\Xi}$ is analogous to Eq.~(5) in \cite{2019MNRAS.487.1047M}, where the inhomogeneous reionization process is captured in large-scale boxes by a semi-analytical {\sc 21cmFAST} simulation \citep{2007ApJ...669..663M,2011MNRAS.411..955M} while the small-scale gas dynamics and baryon structure changes in reaction to the ionization fronts are extracted from a small box hydrodynamic {\sc Gadget-2} simulation \citep{2022MNRAS.513..117L}. The cross-power spectrum is defined as
\begin{eqnarray}
&& \!\!\!\!\!\!\!\!\!\!\!\!\!\!\!\!\!\!\!\!\!\!\!\!
(2\pi)^3 \delta^{(3)}(\textbf{k}-\textbf{k}') P_{\rm m,\Xi}(z_{\rm obs},k)
\nonumber \\
& =& \int_{\mathbb{R}^3} d^3 \boldsymbol{r'} e^{-i\boldsymbol{k}\cdot\boldsymbol{r'}}\langle \Tilde{\delta}_m^{*}(z_{\rm obs},\boldsymbol{k})\Xi(z_{\rm re}(\boldsymbol{r'}),z_{\rm obs}\,|\,\overline{z}_{\rm re})\rangle .
\end{eqnarray}
Analogously to \cite{2019MNRAS.487.1047M}, the cross-power spectrum is written as
\begin{eqnarray} 
    P_{\rm m,\Xi}(k,z_{\rm obs}) \!\!&=&\!\! -\int_{{z_{\rm min}}}^{z_{\rm max}}\frac{\partial \Xi(z_{\rm re},z_{\rm obs})}{\partial z_{\rm re}}P_{\rm m,x_{HI}}(k,z_{\rm re})
    \nonumber \\ && ~~~~\times
    \frac{D(z_{\rm obs})}{D(z_{\rm re})}\,{\rm d}z_{\rm re},
\label{eq:P_mxi}
\end{eqnarray}
where [$z_{\rm min}$=5.5,$z_{\rm max}$=12.0] covers the EoR  and $D(z)$ is the growth function. We extract the cross-power spectrum of matter and neutral hydrogen fraction $P_{\rm m,x_{HI}}$ from {\sc 21cmFAST} simulations, see \S\ref{subsection:21cmFAST} for simulation details.

The noise power spectrum for an interferometer, $P_{\rm N}$ in Eq.~(\ref{eq:P_21}), is given by \citep{2015ApJ...803...21B,2018JCAP...05..004O}
\begin{eqnarray}
    P_{\rm N} (\boldsymbol{k},z) & = & T^2_{\rm sys}(z) \chi^2(z) \lambda (z) \frac{1 + z}{H(z)} \left(\frac{\lambda^2 (z)}{A_e}\right)^2 \left(\frac{S_{\rm area}}{{\rm FOV}(z)}\right) \nonumber \\
    \label{eq:p_thermal}
    & & \times \left(\frac{1}{N_{\rm pol} t_{\rm int} n_{\rm b}(u = k_{\perp} \chi (z)/ 2\pi)}\right) \, ,
\end{eqnarray}
where $T_{\rm sys}$ is the system temperature of the instrument, $\chi$ is the comoving distance, $\lambda (z) = 21 \, {\rm cm} \, (1 + z)$ is the observed wavelength, $S_{\rm area}$ is the area covered by the survey, $N_{\rm pol} = 2$ is the number of polarizations per station feed, and $t_{\rm int}$ is the total integration time. The effective collecting area per station is $A_e = \pi (D_{\rm eff} / 2)^2 = \pi (\sqrt{0.7} D / 2)^2$ where $D$ is the diameter of the receiver and the factor of $0.7$ arises from the assumption that the effective dish area is a scaled version of the physical dish area with an aperture efficiency of seventy per cent. The field of view is then ${\rm FOV}(z) = (\lambda(z) / D_{\rm eff})^2$. The number density of baselines $n_{\rm b}$ in the uv-plane (averaged over a 24 hour period to account for the sky rotation) quantifies the amount of baselines capable of observing a given wavenumber. 

Throughout this work we consider two 21 cm radio telescopes, SKA1-LOW \citep[e.g. ][]{2019arXiv191212699B,2015JCAP...03..034V} and PUMA \citep{2018arXiv181009572C,2019BAAS...51g..53S}. For SKA1-LOW, we follow \citealt{2020PASA...37....7S,2022ApJ...933..236Z} and focus on the dense core of the array configuration, which is composed of 224 stations each with 256 dipole antennas distributed on a radius of $\sim 500$ m.\footnote{We highlight that the outer stations are essential for angular resolution and both calibration and foreground removal purposes.} In the case of SKA1-LOW, we choose to use the density of baselines from \cite{2015JCAP...03..034V} (see their Figure 6) and normalize it such that the number density of baselines recovers the total amount of baselines covered in the current configuration of the SKA core. Mathematically: 
\begin{eqnarray}
    \label{eq:norm}
    \int 2 \pi u n_{\rm b}(u) du = N_{\rm b} \frac{N_{\rm b} - 1}{2} .
\end{eqnarray}
Note that Eq.~(\ref{eq:norm}) guarantees that only unique baselines are counted.

For PUMA, $n_{\rm b}$ is taken from Appendix D of \citet{2018arXiv181009572C} where the observational array is approximated by a fitting formula that traces the number of baselines as a function of the physical distance of the antennas, which are distributed in a hexagonal close-packing in a compact circle. This approximation is then transformed into an uv-plane number density and calibrated using Eq.~(\ref{eq:norm}).

The system temperature, for the radio interferometers of interest, is given by the sum of the sky temperature and the temperature of the receiver (including ground reflections), i.e. 
\begin{equation}
T_{\rm sys}(\nu) \ [\textup{K}] =  \begin{cases} 60  \left(\frac{300 \ \textup{MHz}}{\nu} \right)^{2.55} \times 1.1 + 40, & \textup{SKA}\\
25 \left(\frac{400 \ \textup{MHz}}{\nu} \right)^{2.75} + 2.7 + \frac{300}{9}+ \frac{50}{0.81}, & \textup{PUMA}
\end{cases}~.
\end{equation}

The effective collecting area per station of an aperture array, i.e. for SKA1-LOW, needs a slight correction due to its frequency-dependent nature
\begin{equation}
    \label{eq:sk_A}
    A_e (\nu) =  A_{e, \rm{crit}} \times \begin{cases} \left(\frac{\nu_{\rm crit}}{\nu}\right)^2, & \nu > \nu_{\rm crit} \\
    1, & \nu \leq \nu_{\rm crit}
    \end{cases}~,
\end{equation}
where $A_{e, \rm{crit}} = 3.2 \times 256 = 819.2 $ m$^2$ is the collecting area for the 256 dipole antennas of 3.2 m$^2$ and $\nu_{\rm crit} = 110$ MHz \citep{2020PASA...37....7S}.

In Table \ref{tab:pu-sk}, we have tabulated the primary ingredients of Eq.~(\ref{eq:p_thermal}) for the redshift bins of interest. We emphasize that these are representative instrument parameters, and that the final design of these next-generation radio arrays may be somewhat different.

We do not include the impact of ultra-violet background (UVB) fluctuation on the \HI\ distribution. This effect has been studied by the analytical framework in \citet{2014PhRvD..89h3010P}, \citet{2014MNRAS.442..187G}, and \citet{2019MNRAS.482.4777M}. The UVB fluctuation effect should induce similar order of magnitude fluctuation on the \HI\ power spectrum to the patchy reionization effect in this work on the scales of interest \citep{2014PhRvD..89h3010P,2019MNRAS.485.5059U,2022arXiv220907019L}. 

\begin{table*}
\centering
\caption{Description of radio interferometers considered in this work. We consider only the dense core of the array for SKA1-LOW. Note that neglecting the long spiral arms will not significantly change the sensitivity to the 21 cm power spectrum. Our chosen number of receivers has been chosen to match observing mode 1, described in Table 2 of the \href{https://www.skao.int/sites/default/files/documents/d17-SKA-TEL-SKO-0000557_01_-DesignConstraints-1.pdf}{SKA1-LOW Configuration - Constraints \& Performance Analysis} document. Note that this observing mode has a total bandwidth of 300 MHz, which is well-fitted for our redshift range of interest. Likewise, we chose our proposed survey area based on the Deep SKA1-LOW  survey described in \citealt{2020PASA...37....7S}.}
\label{tab:pu-sk}
\begin{tabular}{lcccccccccc}
\hline\hline
Telescope & \multicolumn{5}{c}{SKA1-LOW} & \multicolumn{5}{c}{PUMA} \\
\hline
Redshift range & \multicolumn{10}{c}{$3.5 < z < 5.5$} \\
Observing time ($t_{\rm int})$ & \multicolumn{5}{c}{$1000$ h} & \multicolumn{5}{c}{$1000$ h \& $5$ yr} \\
Sky coverage ($f_{\rm sky}$) & \multicolumn{5}{c}{0.0024  (100 deg$^2$)} & \multicolumn{5}{c}{$0.5$}\\
Dish/station diameter ($D_{\rm phys}$) & \multicolumn{5}{c}{40 m} & \multicolumn{5}{c}{6 m}\\
Maximum baseline ($b_{\rm max}$) & \multicolumn{5}{c}{$\approx 1$ km} & \multicolumn{5}{c}{$\approx 1.5$ km}\\
Number of receivers ($N_b$) & \multicolumn{5}{c}{224 stations} & \multicolumn{5}{c}{32000 antennas}\\
{} & $z = 3.5$ & $z = 4.0$ & $z = 4.5$ & $z = 5.0$ & $z = 5.5$ & $z = 3.5$ & $z = 4.0$ & $z = 4.5$ & $z = 5.0$ & $z = 5.5$ \\
\cmidrule(lr){2-6} \cmidrule(rl){7-11}
System temperature ($T_{\rm sys}$) [K] & 97.1 & 115 & 135 & 159 & 186 & 145 & 161 & 180 & 202 & 227 \\
Field of view (${\rm FOV}$) [deg$^2$] & 2.62 & 3.23 & 3.91 & 4.65 & 5.46 & 116 & 144 & 174 & 207 & 243\\
Eff. coll. area per station ($A_e$) [m$^2$]  & 98.4 & 121 & 147 & 175 & 205 & \multicolumn{5}{c}{19.8}\\
\hline\hline
\end{tabular}
\end{table*}

\subsection{Fisher matrix formalism}

In this work we aim to study the impact of inhomogeneous reionization on the cosmological parameters inference by post-reionization 21 cm intensity mapping. We estimate the parameter shifts due to this effect under the framework of Fisher matrix. For 21 cm power spectrum data vector in $z$-, $k$- and $\mu$-bins $P_{21}(z,k,\mu)$, the covariance is 
\begin{equation}
    C(z,k,\mu) = \frac{(P_{21}^{\rm fid}(z,k,\mu))^2}{N_{\rm mode}}=\frac{(P_{21}^{\rm fid}(z,k,\mu))^2}{(2\pi)^{-2}\,{\rm d}V_{\rm sur}(z)k^2\,{\rm d}k\,{\rm d}\mu},
\end{equation}
where $V_{\rm sur}$ is the comoving survey volume. The cosmological parameters involved in our Fisher matrix include: today's Hubble constant $h=H_0/(100\,\rm km\,s^{-1}\,Mpc^{-1})$, the baryon density parameter $\Omega_{\rm b}h^2$,  the cold dark matter density parameter $\Omega_{\rm c}h^2$, the Thomson optical depth to reionization $\tau$, the sum of neutrino masses $\sum m_{\nu}$, the scalar spectral index $n_s$, the primordial amplitude $A_{\rm s}$, \HI\ bias coefficient $b_{\rm HI}$, and neutral hydrogen fraction $\Omega_{\rm HI}$ (in each redshift bin). The fiducial values of them are summarized in Table. \ref{tb:params}. For the sake of avoiding numerical complication in the Fisher matrix calculation, the actual parameter vector in the code is rescaled as $\Vec{p}=\{h,\,\Omega_{\rm b}h^2$, $\Omega_{\rm c}h^2$, $\tau$, $\sum m_{\nu}$, $n_{\rm s},\, 10^9A_{\rm s},\,b_{\rm HI},\,10^3\Omega_{\rm HI}\}$.

\begin{table}
\caption{A summary of cosmological and astrophysical parameters, and their fiducial values. For $b_{\rm HI}$ we cite values in \citet{2018ApJ...866..135V} and for $10^3\Omega_{\rm HI}$ we use the results measured in \citet{2015MNRAS.452..217C}. }
\begin{tabular}{ccc}
\hline
\hline
Parameter &  \multicolumn{2}{c}{Fiducial}  \\\hline
$ h$ & \multicolumn{2}{c}{0.6774} \\
 $\Omega_{\rm b}h^2$ & \multicolumn{2}{c}{0.0223} \\
 $\Omega_{\rm c}h^2$ & \multicolumn{2}{c}{0.1188}\\
 $\sum m_{\nu}$ [eV] & \multicolumn{2}{c}{0.194} \\
 $10^9 A_s$ & \multicolumn{2}{c}{2.142} \\
 $n_{\rm s}$ & \multicolumn{2}{c}{0.9667} \\
 $\tau$ & \multicolumn{2}{c}{0.066}\\
 \hline
~ & $3.5<z<4.5$ & $4.5<z<5.5$ \\
$b_{\rm HI}$ & 2.82 & 3.18  \\
$10^3\Omega_{\rm HI}$ & 1.18 & 0.98 
\\
\hline
\hline
\end{tabular}\label{tb:params}
\end{table}

Then the Fisher matrix element for parameters $l$ and $m$ is
\begin{eqnarray}
    F_{lm} \!\!\!\!
    &= & \!\!\!\!\frac{1}{2}\sum_{z,k,\mu} \frac{\partial P_{21}^{\rm fid}(z,k,\mu)}{\partial p_l}C^{-1}(z,k,\mu)\frac{\partial P_{21}^{\rm fid}(z,k,\mu)}{\partial p_m} \nonumber \\
    &= &\!\!\!\!\frac{1}{8\pi}\int^{z_{\rm max,bin}}_{z_{\rm min,bin}}{\rm d}V_{\rm sur}(z)\int^{+1}_{-1}{\rm d}\mu \int^{k_{\rm max}}_{k_{\rm min}} k^2 {\rm d}k  \nonumber \\
    &&~ \times \frac{\partial P_{21}^{\rm fid}(z,k,\mu)}{\partial p_l}\frac{1}{(P_{21}^{\rm fid}(z,k,\mu))^2}\frac{\partial P_{21}^{\rm fid}(z,k,\mu)}{\partial p_m}. ~~~~~~
\end{eqnarray}
In this work, we have 10 redshift bins spanning $3.5<z<5.5$ with $\Delta z =0.2$, the survey volume in each redshift bin is calculated as
\begin{equation}
    V_{\rm sur}= \frac{4\pi}3f_{\rm sky}[D_{\rm C}^3(z_{\rm max,bin}) - D_{\rm C}^3(z_{\rm min,bin})].
\end{equation}

 We choose $k_{\rm max}=0.4 \ \textup{Mpc}^{-1}$ as the linearity cut-off, the variance per ${\rm d}\ln k$ of the matter field $\Delta(k_{\rm max}) = [k_{\rm max}^3P_{\rm m}(k_{\rm max}, z)/2\pi^2]^{1/2}=0.38$ at $z=3.5$, and that of the galaxies is 1.07. The large-scale cutoff $k_{\parallel, \rm min}$ is set by requiring one wavelength across the redshift shell we are using,
\begin{equation}
    k_{\parallel \rm,min} = \frac{2\pi}{D_{\rm C}(z_{\rm max,bin})-D_{\rm C}(z_{\rm min,bin})}.
\end{equation}
The minimum cosmological transverse wavenumber $k_{\perp,\rm min}$ being sampled is  
\begin{equation}
    k_{\perp,\rm min}=\frac{2\pi D_{\rm phys}}{\lambda_{\rm obs}D_{\rm C}(z)}
  \end{equation}
where $D_{\rm phys}$ is the distance between two antennas, we use the value of dish diameter in our calculation, $\lambda_{\rm obs}=\lambda_{21}(1+z)$ is the observed 21 cm signal wavelength. 

For cosmological 21 cm experiment, the foreground contamination signal from synchrotron and free-free emission could be several order of magnitude larger than that from IGM \HI. Nonetheless, these astrophysical foregrounds are spectrally smooth and only enter into small $k_{\parallel}$ Fourier modes. Precisely, the contamination mostly affect wavenumbers in a ``wedge'' \citep{2010ApJ...724..526D, 2012ApJ...752..137M, 2012ApJ...745..176V, 2016MNRAS.456.3142S} in Fourier space, where

\begin{eqnarray}
    \mu\leq \mu_{\rm min} \!\!\!\!
    &=& \!\!\!\! \frac{k_{\parallel,\rm wedge\, min}}{\sqrt{k_{\parallel,\rm wedge\, min}^2+k_{\perp,\rm min}^2}}\\ \notag
    \!\!\!\! &=& \!\!\!\! \frac{D_{\rm C}(z)H(z)/[c(1+z)]}{\sqrt{1+\{D_{\rm C}(z)H(z)/[c(1+z)]\}^2}}
\end{eqnarray}

In this work, we present the results in \S~\ref{sec:results} aligning with the optimistic forecasting choice by PUMA white paper \citep{2018arXiv181009572C} that this technical problem will be overcome by instrumental designs and methodological advances such that full wedge calibration could be rendered when the 21 cm data is to be collected. 
We show the cosmological parameters shifts when the ``wedge'' modes cannot be recovered in Table~\ref{table:shifts_wdgon} in Appendix~\ref{appendix:wdgoff}. When computing Fisher matrix, we have 30 logarithmically spaced $k$-bins from $k_{\rm min}=\sqrt{k_{\parallel,\rm min}^2+k_{\perp,\rm min}^2}$ to $k_{\rm max}$, 10 equally sliced $\mu$-bins where $|\mu| > \mu_{\rm min}$ for each $k$ bin when the ``wedge'' issue is taken into account. We show the values of the $\mu_{\rm min}$ of wedge effect, $k_{\parallel,\rm min}$, and $k_{\perp,\rm min}$ for both PUMA and SKA1-LOW in each redshift bin in Table~\ref{tab:modes}.

\begin{table}
\centering
\caption{Wedge angle $\mu_{\rm min}$, $k_{\parallel,\rm min}$ and $k_{\perp,\rm min}$ for each redshift bin.}
\label{tab:modes}
\begin{tabular}{ccccc}
\hline\hline
z & $\mu_{\rm min}$ & $k_{\parallel,\rm min}$ & \multicolumn2c{$k_{\perp,\rm min}$}
\\ \hline
 & & & PUMA & SKA1-LOW 
 
\\
\cmidrule(lr){4-5}
3.6 & 0.89 & 0.0392 & 0.0055 & 0.0369 
\\
3.8 & 0.89 & 0.0418 & 0.0052 & 0.0346 
\\
4.0 & 0.90 & 0.0444 & 0.0049 & 0.0325  
\\
4.2 & 0.91 & 0.0470 & 0.0046 & 0.0307 
\\
4.4 & 0.91 & 0.0497 & 0.0044 & 0.0291 
\\
4.6 & 0.92 & 0.0525 & 0.0041 & 0.0276 
\\
4.8 & 0.92 & 0.0553 & 0.0039 & 0.0262 
\\
5.0 & 0.93 & 0.0581 & 0.0038 & 0.0250 
\\
5.2 & 0.93 & 0.0610 & 0.0036 & 0.0239 
\\
5.4 & 0.93 & 0.0640 & 0.0034 & 0.0229 
\\
\cmnt{
3.6 & 0.0392 & 0.0369 & 0.89 
\\
3.8 & 0.0418 & 0.0346 & 0.89 
\\
4.0 & 0.0444 & 0.0325 & 0.90 
\\
4.2 & 0.0470 & 0.0307 & 0.91 
\\
4.4 & 0.0497 & 0.0291 & 0.91 
\\
4.6 & 0.0525 & 0.0276 & 0.92 
\\
4.8 & 0.0553 & 0.0262 & 0.92 
\\
5.0 & 0.0581 & 0.0250 & 0.93 
\\
5.2 & 0.0610 & 0.0239 & 0.93 
\\
5.4 & 0.0640 & 0.0229 & 0.93 
\\
}
\hline\hline
\end{tabular}
\end{table}

Our first 7 parameters are constant through the redshifts $3.5<z<5.5$ we consider in this work. However we allow different values of $b_{\rm HI}$ and $10^3\Omega_{\rm HI}$ in $3.5<z <4.5$ and $4.5< z <5.5$ to allow for redshift evolution between the two bins (see Table \ref{tb:params} about the fiducial values of all parameter). To implement this, we separate the parameter vectors into a subspace that does not depend on redshift $\Vec{g}=\{h,\, \Omega_{\rm b}h^2$,\, $\Omega_{\rm c}h^2$,\, $\tau$,\, $\sum m_{\nu}$,\, $n_{\rm s},\, 10^9A_{\rm s}\}$, and a subspace that takes on a different value in each bin $\Vec{s}=\{b_{\rm HI},10^3\Omega_{\rm HI}\}$. Then:
\newcommand{\bigzero}{\mbox{\normalfont\Large\bfseries 0}}
\begin{equation}
{\mathbfss F} =
\left(
\begin{array}{c|cc}
{\mathbfss F}_{\rm gg} &  {\mathbfss F}_{\rm gs,\,z-bin\,1} & {\mathbfss F}_{\rm gs,\,z-bin\,2}\\\hline
\\
{\mathbfss F}_{\rm sg,\,z-bin\,1} & {\mathbfss F}_{\rm ss,\,z-bin\,1} & \mathbb{O} \\
\\
{\mathbfss F}_{\rm sg,\,z-bin\,2} & \mathbb{O} & {\mathbfss F}_{\rm ss,\,z-bin\,2}
\end{array}
\right),
\end{equation}
where ${\mathbfss F}_{\rm gg}$ sums over the full range $3.5\le z\le 5.5$, the ``z-bin~1'' submatrices sum over only $3.5\le z\le 4.5$, and the ``z-bin 2'' submatrices sum over $4.5\le z\le 5.5$.

\subsection{Priors} 

To break degeneracy of certain parameters (mostly $\Omega_{\rm HI}$ and $A_{\rm s}$ in our case) in the information matrix, we add priors on $\{\Omega_{\rm b}h^2, \Omega_{\rm c}h^2, \tau, \sum m_{\nu}, h, A_{\rm s}\}$ from \textit{Planck} 2018 ‘base + $m_{\nu}$’ parameter chains with ‘TT+TE+lowl+lowE+lensing’ data combination \citep{2020A&A...641A...6P} released through the Planck Legacy Archive (\href{https://pla.esac.esa.int}{PLA})\footnote{\href{https://pla.esac.esa.int}{https://pla.esac.esa.int}}. This is a ``minimal CMB'' prior, as much more CMB data will be available at the time of an ambitious Stage II 21 cm IM experiment; but it prevents us from having to internally constrain all of these parameters from shape information in the matter (hence 21 cm) power spectrum over a modest redshift range.

In the Fisher matrix formalism, the parameter shift due to patchy reionization can be expressed using the general formula for shifts due to the linear effects of systematics \citep{1998PhRvL..81.2004K,2020JCAP...04..006L}:
\begin{equation}
    \Delta p_l =  \sum_{z,k,\mu}\sum_{m}F^{-1}_{lm} \frac{\partial P_{21}^{\rm fid}(z,k,\mu)}{\partial p_m}C^{-1}(z,k,\mu)P_{21}^{\rm patchy}(z,k,\mu) .
\end{equation}

\section{Simulations}\label{sec:sims}
Modeling the effect of inhomogeneous reionization in the post-reionization IGM and its subsequent impact on 21 cm line intensity mapping requires a \emph{large} dynamical range due to the physical scales of interest. First, one requires the mass resolution necessary to carefully track how structures near the filtering length react to the passage of ionization fronts \citep{2022MNRAS.513..117L}. Simultaneously, the patchy nature of reionization, which couples to the reionization bubbles scale, must be accounted for. Here we adopt an analog hybrid strategy to the one used in \cite{2019MNRAS.487.1047M,2020MNRAS.499.1640M} where a semi-numerical simulation with large box side capable of modeling the patchy nature of reionization is used jointly with a small box side hydro-dynamical simulation that carefully tracks the way gas reionizes.

\begin{figure}
    \centering
    \includegraphics[width=\linewidth]{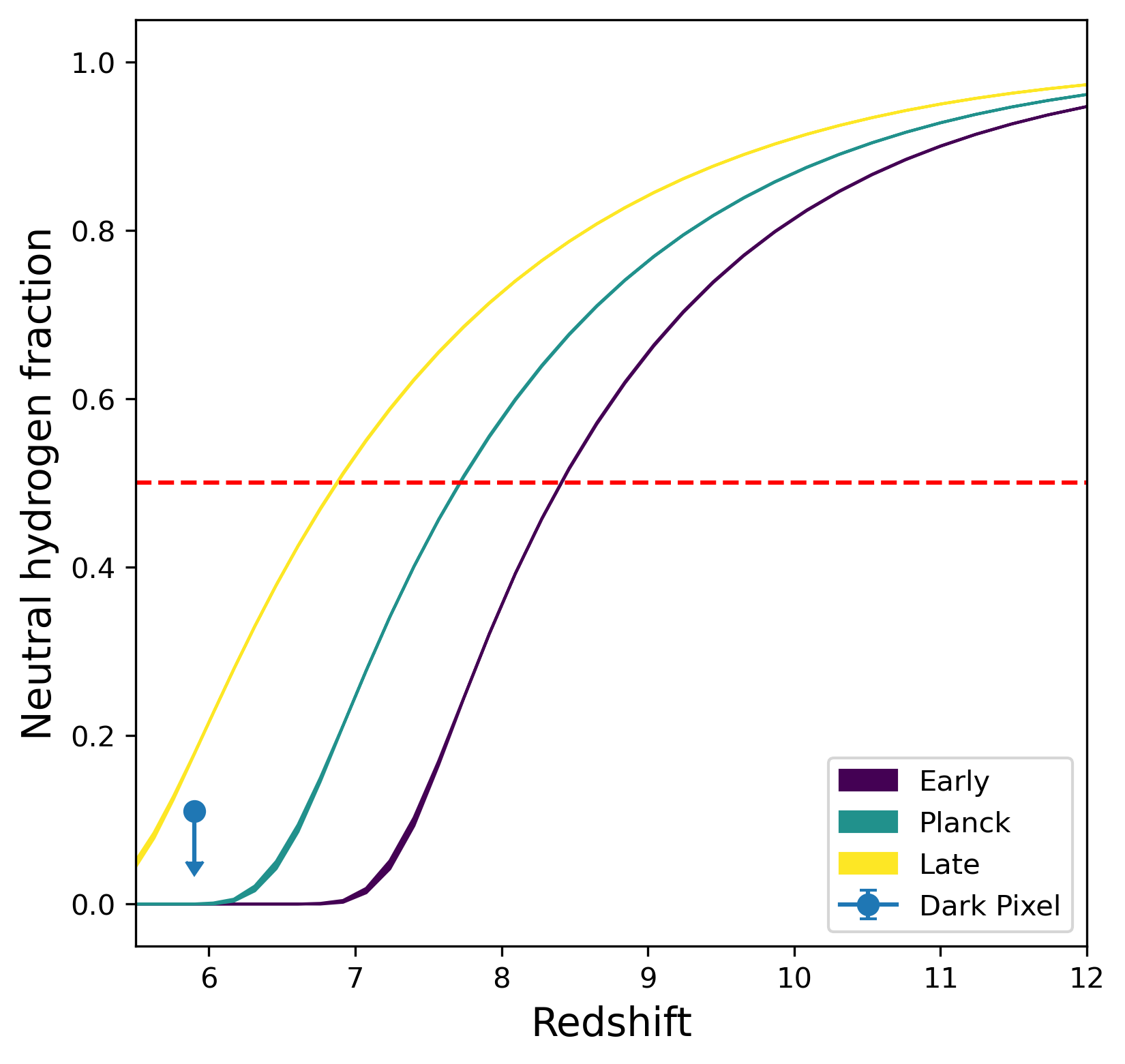}
    \caption{Reionization history for the three models considered throughout our work. The red dashed line corresponds to the mid-point of reionization. We show the dark pixel constraint \citep{2015MNRAS.447..499M} as the blue dot near $z\approx 5.9$. Error bands due to Monte Carlo scatter from the 4 realizations are included although they can only be appreciated at the tail end of the reionization process -- around $z\approx 7.2$ for the early reionization model.}
    \label{fig:history}
\end{figure}

\subsection{{\sc 21cmFAST} simulation suite} \label{subsection:21cmFAST}

The \textit{large box} simulations take into account the inhomogeneous nature of the reionization  process. The main purpose of these simulations is to study how the matter density is correlated with the ionized bubble formation and its subsequent evolution, i.e. to obtain $P_{\rm m, x_{HI}}$ for Eq.~(\ref{eq:P_mxi}). 

We use {\sc 21cmFAST} -- version $3.0.4$ -- for our large box simulations \citep{2011MNRAS.411..955M, 2020JOSS....5.2582M}. The box side is 400 Mpc comoving, with $256^3$ cells for \ion{H}{I} and $768^3$ cells for the matter field. We mostly use {\sc 21cmFAST} default parameters and set the cosmology to match that of \citet{2016A&A...594A..13P}. In the default setting of {\sc 21cmFAST}, Population II stars drive the reionization process, therefore we change the number of ionizing photons per baryons of Population II stars to study the impact of different reionization histories in 21 cm line intensity mapping in the post-reionization era. Specifically, we use three models: fiducial model, which has a mid-point of reionization $z_{\rm mid} = 7.74$ consistent with the reionization history inferred from \textit{Planck} 2018 `TT+TE+EE+lowE+lensing' \citep{2020A&A...641A...6P} and early (late) reionization model with $z_{\rm mid} = 8.46$ ($z_{\rm mid} = 6.92$), which are consistent with the 1$\sigma$ upper (lower) limit from the Planck's results. Although these three models allow us to establish the preliminary dependence of the memory of reionization with the timing of reionization, they do not fully sample the distribution of observationally allowed reionization models \citep{2021MNRAS.504.1555M}. Future work will further investigate the dependence of the memory of reionization in 21 cm intensity mapping with cosmic dawn and reionization. We put the data generated using {\sc 21cmFAST} in a GitHub repository \footnote{\href{https://github.com/CosmoSheep/reion\_ns}{https://github.com/CosmoSheep/reion\_ns}}.

We plot the different reionization histories considered throughout this work in Figure~\ref{fig:history}. Note that the late reionization scenario is in slight tension with the upper limit constraint obtained from dark pixel measurements in the \lya forest \citep{2015MNRAS.447..499M}. However, large \lya opacity fluctuations (consistent with those expected in the end stages of reionization) have been reported at $z \sim 5.5$ \citep{2015MNRAS.447.3402B}. Naturally, models consistent with these observations have \emph{ultra}-late reionization scenarios where neutral hydrogen islands are present even at $z = 5.5$ \citep[e.g. ][]{2020MNRAS.491.1736K, 2019MNRAS.485L..24K, 2020MNRAS.494.3080N}. Nevertheless, the distribution of these low-transmission regions depends on the ionization level of voids, and thus correlates strongly with the spectrum above the gap detection threshold, which implies that these dark gaps by themselves may not be sufficient to constrain the end stage of reionization \citep{2022ApJ...937...17G}. 

\begin{figure}
    \centering
    \includegraphics[width=\linewidth]{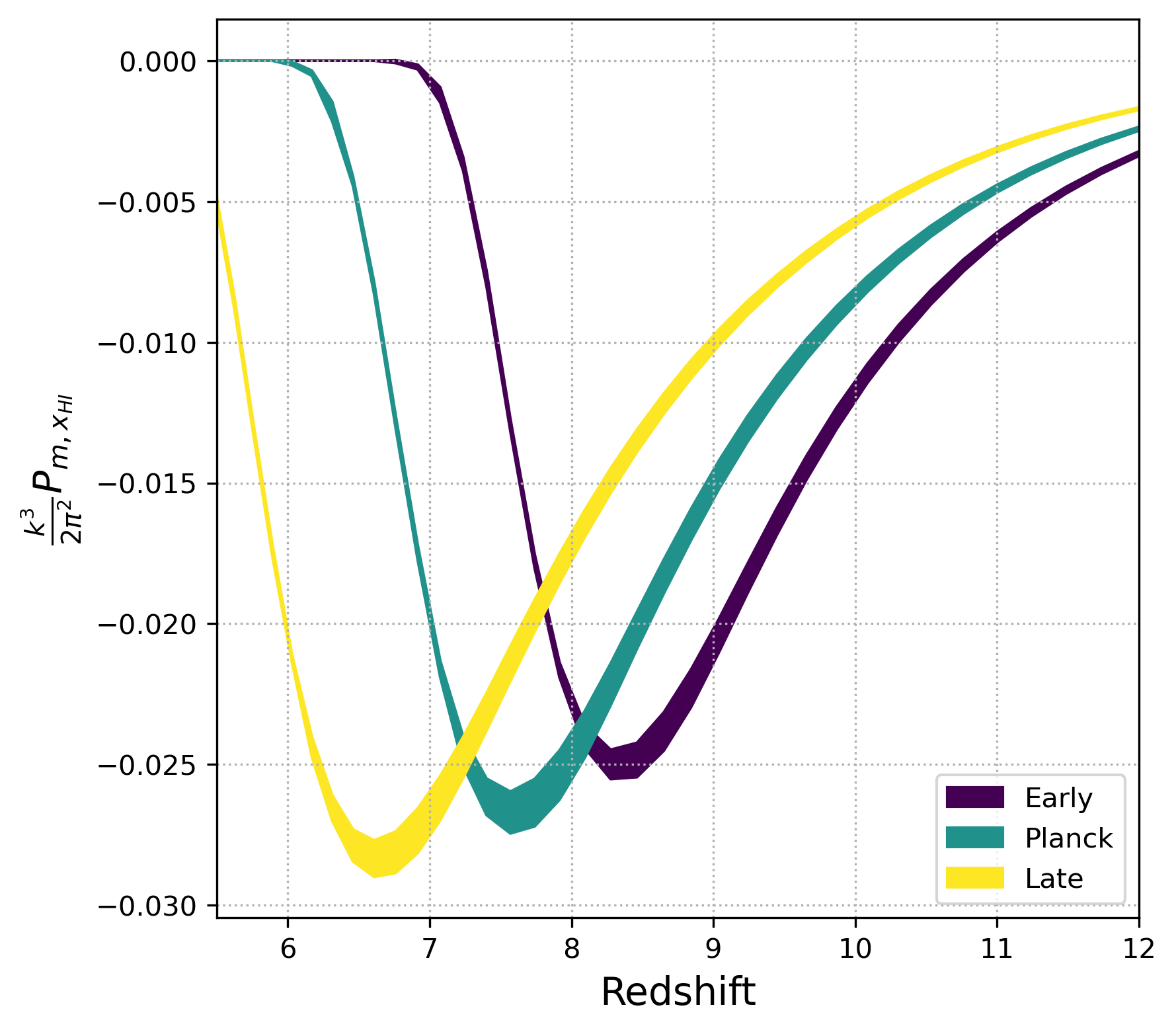}
    \caption{The cross-power spectrum of matter and neutral hydrogen fraction as a function of redshift and evaluated at $k = 0.14$ Mpc$^{-1}$. Bands show the Monte Carlo scatter from 4 realizations.}
    \label{fig:cross_ps}
\end{figure}

For each model we run four different realizations in order to reduce the variance of the simulations. In each run, we obtain a total of 92 snapshots that trace both the density and neutral hydrogen fields in the range of $5 \leq z \leq 35$, with a step size of 2\% in scale factor, i.e., $1 + z_{i+1} = 1.02(1+z_i)$.
With these snapshots we compute the cross-power spectrum of matter and neutral hydrogen fraction $P_{\rm m,x_{HI}}$, which reflects the effect of patchy reionization in the IGM since it correlates how matter and ionized bubble structure are distributed. We plot the dimensionless cross-power spectrum of matter and neutral hydrogen fraction for all three models as a function of redshift and evaluated at $k = 0.14$ Mpc$^{-1}$ in Figure~\ref{fig:cross_ps}.

Figure~\ref{fig:cross_ps} shows that matter and neutral hydrogen fields are anti-correlated because reionization happens ``inside-out'' -- dense regions that host and surround the sources of ultraviolet photons ionize first. Furthermore, $P_{\rm m,x_{HI}}$ peaks (in absolute value) near the midpoint of reionization for a given model. Note that later reionization scenarios lead to a stronger cross-power spectrum of matter and neutral hydrogen fractions and consequently a stronger memory of reionization  on 21 cm intensity mapping in the post reionization era. Naturally, earlier reionization leads to a slightly lesser signal since our ``Early'' model only changes the overall abundance of ionizing photons. Thus, more available UV photons will lead to a slightly faster reionization timeline (see Figure \ref{fig:history}).  

Due to the box size of our {\sc 21cmFAST} simulations (400 Mpc), our snapshots do not have enough modes to reliably extract the cross-power spectrum of matter and neutral hydrogen fraction for $k_{\rm cut} \lessapprox 0.053\,{\rm Mpc}^{-1}$. For instance, the first two bins in the box have $\leq 4$ modes each, hence we choose to implement a cutoff based on $k_{\rm cut}$, which corresponds to a bin with 72 modes. To model scales larger than this cutoff in $P_{\rm m,x_{HI}}$, we extrapolate using a linear biasing model,
\begin{equation}
    P_{\rm m, x_{HI}}(k < k_{\rm cut},z) = \frac{P_{\rm m, x_{HI}}(k_{\rm cut},z)}{P_m (k_{\rm cut},z)} P_m (k, z) \, ,
\end{equation}
which should be valid on scales larger than the size of the ionization bubbles since the fundamental assumption of the biasing models is locality.

\subsection{GADGET-2 simulation suite}\label{ssec:gadget}

We used a modified version of {\sc Gadget-2}, as previous implemented and tested in \citet{2018MNRAS.474.2173H,2022MNRAS.513..117L}, to track the reactions of small-scale baryon structures to reionization. The \textit{small box} {\sc Gadget-2} simulation has box size $L=1152{\rm\,h^{-1}\,kpc}$, and number of particles, $N=2\times(256)^3$. Reionization is implemented by resetting the temperature of gas particles to $2\times 10^4$ instantaneously at $z_{\rm re}$. We simulate four realizations to reduce the statistical error due to the
limited box size by a factor of $\sqrt{4}$. We archived our Gadget-2 simulation suite and tabulated filtering mass data in a Github repository\footnote{\href{https://github.com/CosmoSheep/HIPowerSpectrum}{https://github.com/CosmoSheep/HIPowerSpectrum}}.

\section{Results}\label{sec:results}

\subsection{Post-reionization {H\,\sc{i}} abundance}

\begin{figure}
    \centering
    \includegraphics[width=1\columnwidth]{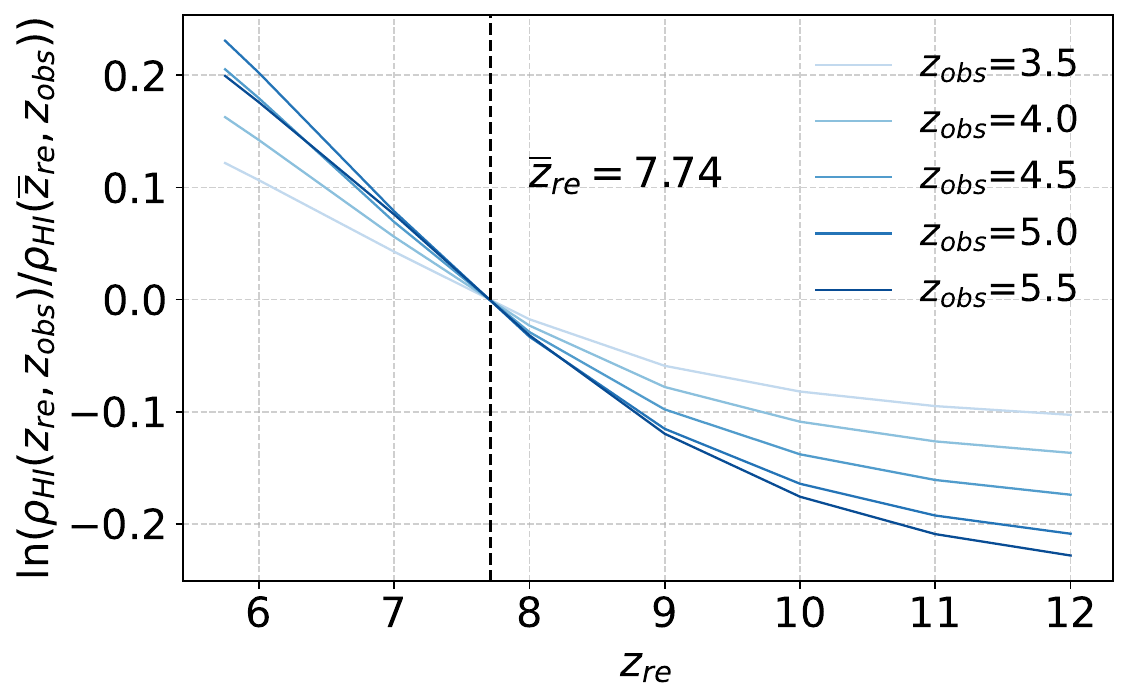}
    \caption{The \HI\ overdensity fluctuation considering different instantaneous reionization time with $\overline{z}_{\rm re}=7.74$. The magnitude of overdensity attenuates through time after reionization but is still not negligible at redshift as low as 3.5.}
    \label{fig:xi}
\end{figure}

We plot in Figure~\ref{fig:xi} the fractional \HI\ fluctuation $\Xi(z_{\rm re},z_{\rm obs})$ due to inhomogeneous reionization defined in Eq.~(\ref{eq:xi}). With $\overline{z}_{\rm re}=7.74$, the fluctuations of \HI\ overdensity spread from approximately $-0.2$ to $+0.2$ for $z_{\rm obs}=5.5$ and $-0.1$ to $+0.1$ for $z_{\rm obs}=3.5$. This magnitude of overdensity fluctuations is non-negligible and leads to a considerable contribution to the 21 cm IM power spectrum, as will be shown later. Note that for the $z_{\rm obs}=5.5$ curve, at the left end when $z_{\rm re}$ approaches 6 the overdensity does not go over that of $z_{\rm obs}= 5.0$ curve. This is because the most violent \HI\ gas relaxation happens right after the end of reionization (Figure 2 of \citealt{2018MNRAS.474.2173H} demonstrates this phenomenon by the comparison of velocity distributions between $z=6.5$ and $z=3.5$). With the observation redshift $z_{\rm obs}=5.5$ very soon to the end of reionization, the relative \HI\ density change in scenarios $6\le z_{\rm re}\le 7$ will not be as distinguishable as those at lower observation redshifts. 

\begin{figure}
    \centering
    \includegraphics[width=1\columnwidth]{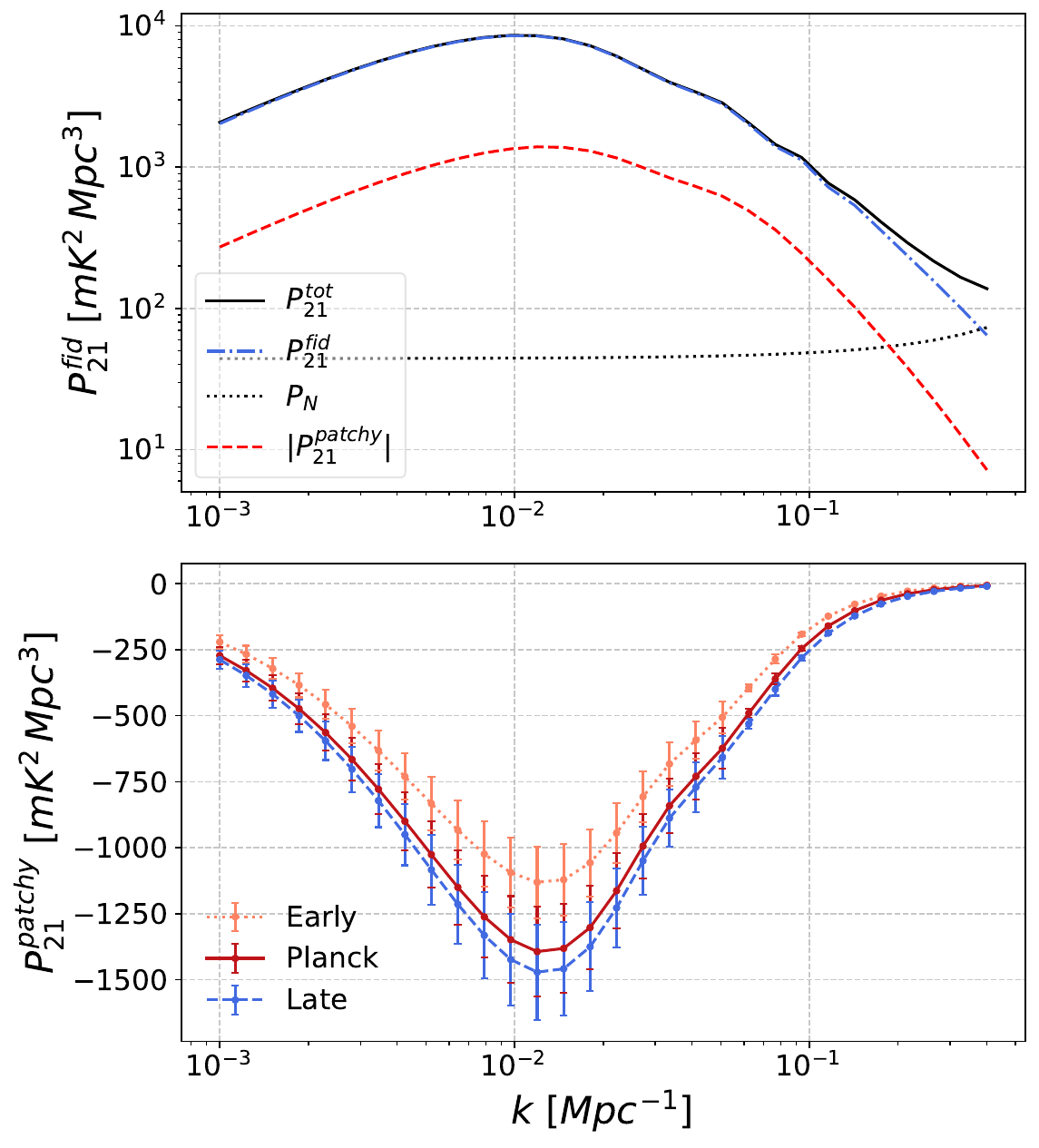}
    \caption{(\textit{Upper panel}) 21 cm power spectrum at $z=5.0$, $\mu=1/\sqrt{3}$. Dotted black line represents thermal noise power spectrum calculated with setup of PUMA and $t_{\rm int}=$ 5 years. The dashed red line represents the absolute value of power spectrum induced by inhomogeneous reionization $|P_{21}^{\rm patchy}|$ in Planck Reionization scenario (for the sake of comparison we flip the sign as $P_{21}^{\rm patchy}$ is negative). The dotted dash blue line is the 21 cm power spectrum of fiducial cosmology $P_{21}^{\rm fid}$ and solid black line is the total 21 cm power spectrum. (\textit{Lower panel}) the patchy reionization power spectrum of three reionization scenarios. The error bars show the Monte Carlo scatter from 4 realizations of {\sc 21cmFAST} simulation (uncertainties from {\sc GADGET-2} realizations are negligible, $\sim 0.1\%$ compared to $\sim 10\%$ from {\sc 21cmFAST}, so we do not display them in figures).}
    \label{fig:ps}
\end{figure}

\begin{figure}
    \centering
    \includegraphics[width=1\columnwidth]{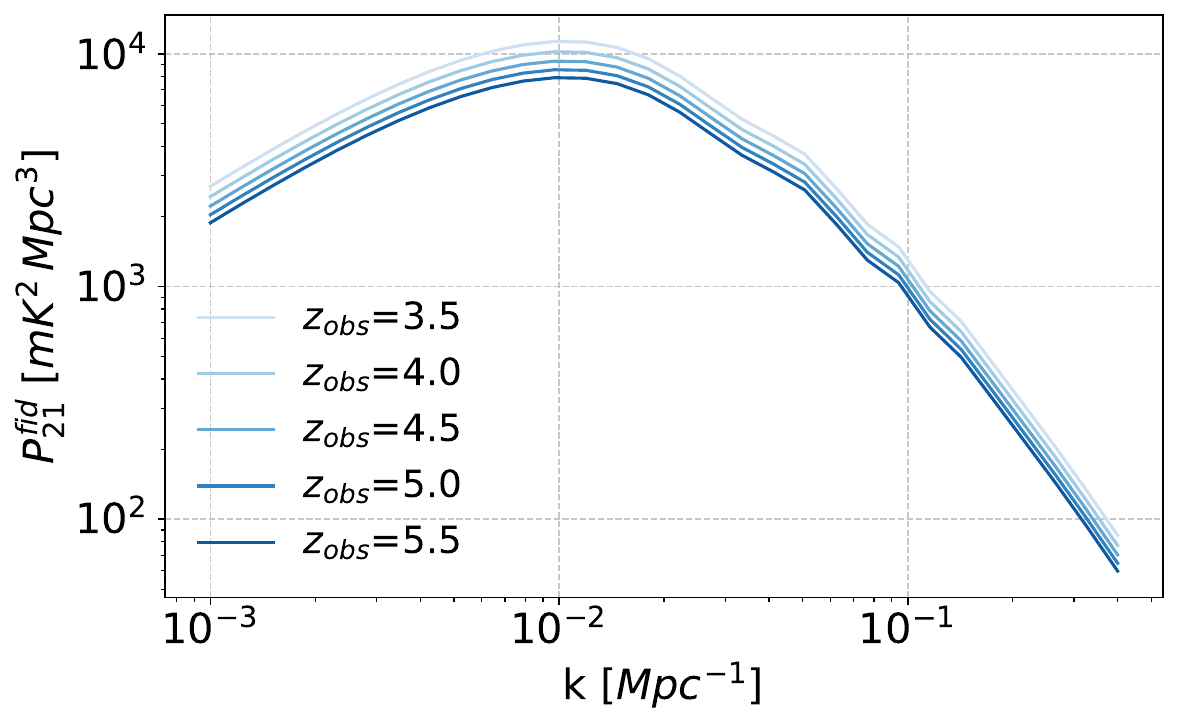}
    \caption{The fiducial 21 cm power spectra with respect to wavenumber k with observation redshift from 3.5 to 5.5. }
    \label{fig:p_fid}
\end{figure}

\begin{figure}
    \centering
    \includegraphics[width=1\columnwidth]{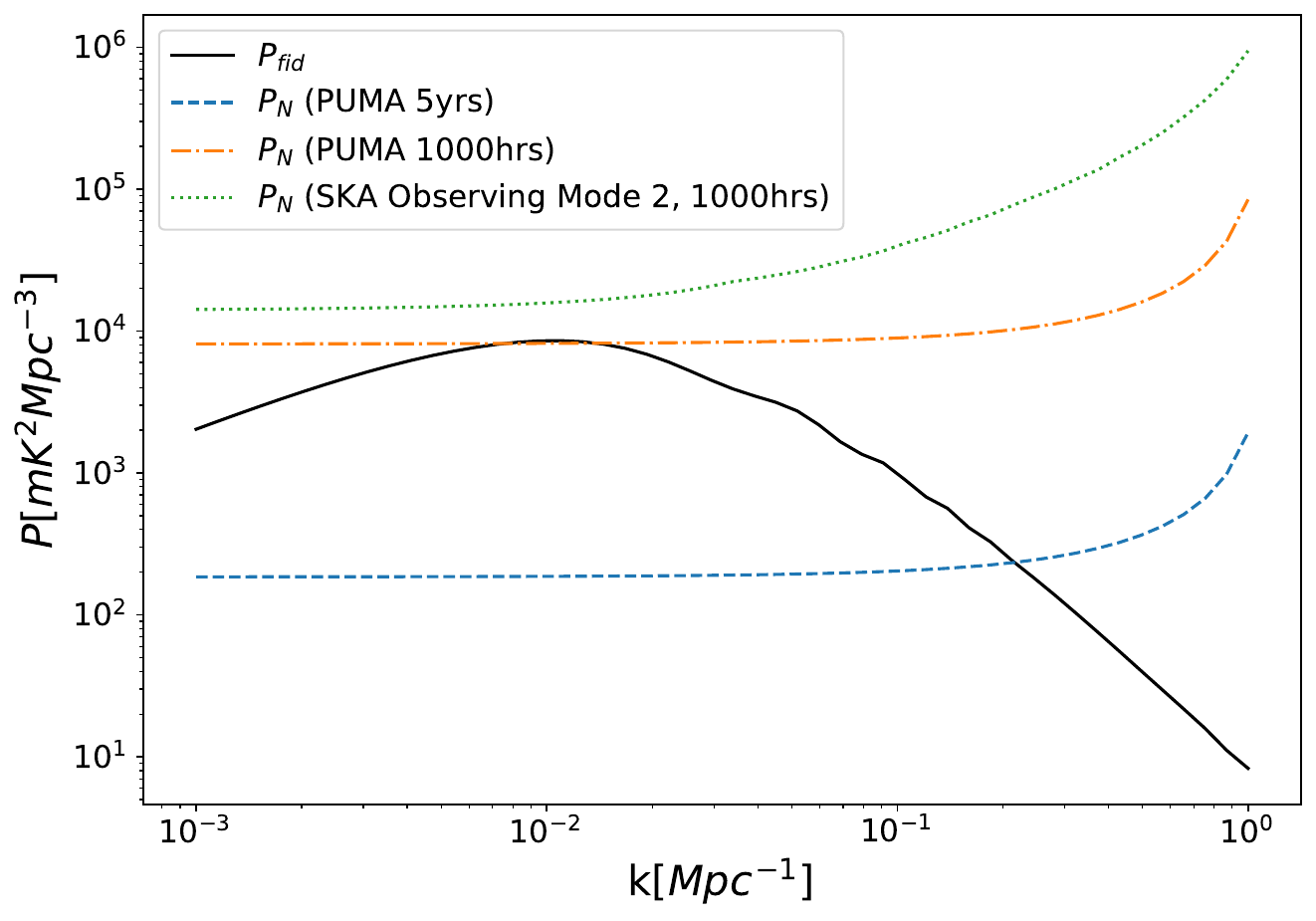}
    \caption{The thermal noise power spectra $P_{\rm N}$ for three obeserving scenarios: PUMA $t_{\rm int}$=5 year, PUMA $t_{\rm int}$=1000 hours and SKA1-LOW PUMA $t_{\rm int}$=1000 hours.}
    \label{fig:p_noise}
\end{figure}

\begin{figure*}
    \centering
    \includegraphics[width=2\columnwidth]{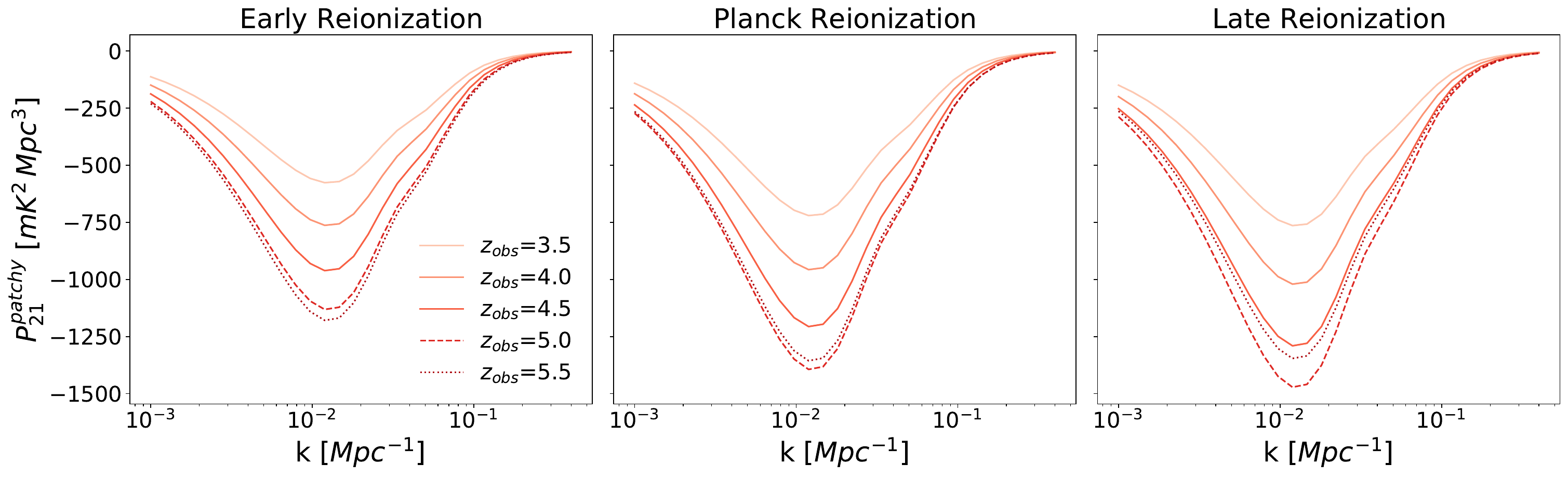}
    \caption{ Power spectra contribution from inhomogeneous reionization $P_{21}^{\rm patchy}$ with respect to wavenumber $k$ of observation redshifts from 3.5 to 5.5 at $\mu=1/\sqrt{3}$. We plot the average values from 4 realization of {\sc 21cmFAST} simulation for each scenario with error bar hidden for the sake of clarity. (\textit{Left panel}) $P_{21}^{\rm patchy}$ calculated in the Early Reionization scenario with $z_{\rm mid}=8.46$. (\textit{Middle panel}) Planck Reionization scenario with $z_{\rm mid}=7.74$. (\textit{Right panel}) Late Reionization scenario with $z_{\rm mid}=6.92$.}
    \label{fig:p_patchy}
\end{figure*}

\begin{figure*}
    \centering
    \includegraphics[width=2\columnwidth]{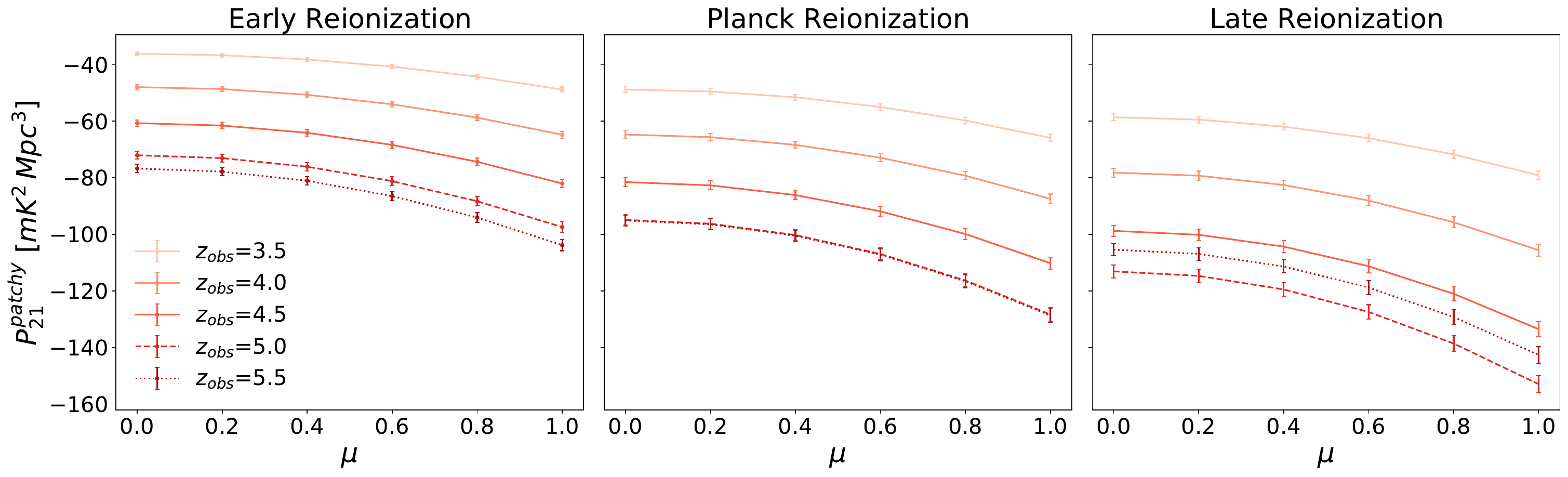}
    \caption{ Power spectra contribution from inhomogeneous reionization $P_{21}^{\rm patchy}$ with respect to $\mu=\cos{\theta}$ of observation redshifts from 3.5 to 5.5 at $k=0.14\,\rm Mpc^{-1}$. The error bars represent the statistical error arised from 4 realizations of {\sc 21cmFAST} simulation. (\textit{Left panel}) $P_{21}^{\rm patchy}$ calculated in the Early Reionization scenario with $z_{\rm mid}=8.46$. (\textit{Middle panel}) Planck Reionization scenario with $z_{\rm mid}=7.74$. (\textit{Right panel}) Late Reionization scenario with $z_{\rm mid}=6.92$.}
    \label{fig:p_patchy_mu}
\end{figure*}

\begin{figure*}
    \centering
    \includegraphics[width=2\columnwidth]{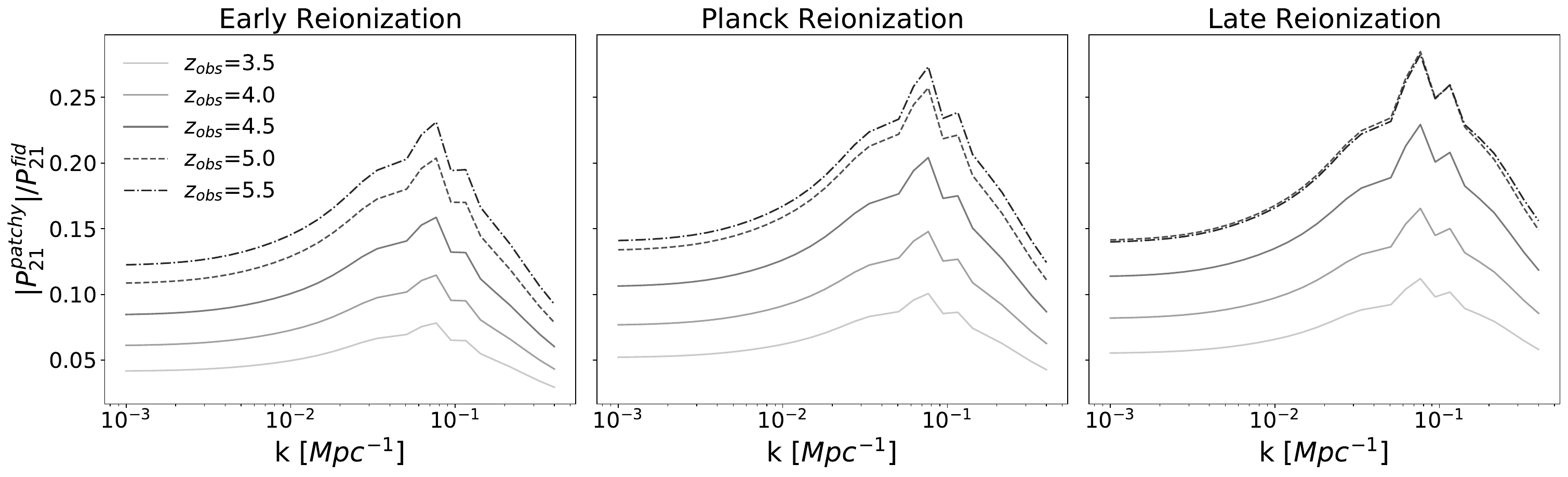}
    \caption{The ratio of absolute value of $P_{21}^{\rm patchy}$ to $P_{21}^{\rm fid}$ of observation redshifts from 3.5 to 5.5 at $\mu=1/\sqrt{3}$, PUMA $t_{\rm int}=5$ years. We plot the average values from 4 realization of {\sc 21cmFAST} simulation for each scenario with error bar hided for the purpose of clarification. With similar shape, the reionization imprints attenuate after EoR. (\textit{Left panel}) $P_{21}^{\rm patchy}$ calculated in the Early Reionization scenario with $z_{\rm mid}=8.46$. (\textit{Middle panel}) Planck Reionization scenario with $z_{\rm mid}=7.74$. (\textit{Right panel}) Late Reionization scenario with $z_{\rm mid}=6.92$.}
    \label{fig:reion_mem}
\end{figure*}

\subsection{21 cm power spectra and reionization memory}

We plot the total 21 cm power spectrum of Eq.~(\ref{eq:P_21}) and contributions from fiducial flashy reionization model, patchy reionization and thermal noise respectively at $z=5.0$ and $\mu=1/\sqrt{3}$ (where $P_2(\mu)=0$) in Figure~\ref{fig:ps}. We plot the fiducial 21 cm power spectra of various $z_{\rm obs}$ in Figure~\ref{fig:p_fid}. We also show the noise power spectra of different observing scenarios from Tabel~\ref{tab:pu-sk} in Figure~\ref{fig:p_noise}.

To look at the impact of reionization on 21 cm power spectrum, we plot in Figure~\ref{fig:p_patchy} the patchy reionization contribution $P_{21}^{\rm patchy}(k,z_{\rm obs})$ of three reionization scenarios with $z_{\rm obs}$ from 3.5 to 5.5. Note that for the Early Reionization ($z_{\rm mid}=8.46$) scenario the $P_{21}^{\rm patchy}$ indeed peaks at $z_{\rm obs}=5.5$ and attenuates afterwards, which is consistent with the intuition that the closer $z_{\rm obs}$ is to the EoR, the larger the imprints of inhomogeneous reionization should be. However, for Planck Reionization ($z_{\rm mid}=7.74$) and Late Reionization ($z_{\rm mid}=6.92$), $P_{21}^{\rm patchy}$ peaks at $z_{\rm obs}=5.0$ instead of 5.5. We interpret these trends as consistent with the rapid gas relaxation immediately following reionization that we observe in Figure \ref{fig:xi}. The calculation of $P_{\rm m, \Xi}(k,z_{\rm obs})$ in Eq.~(\ref{eq:P_mxi}) is sensitive to the relative \HI\ abundance change with respect to $z_{\rm re}$, which is small at the time right after EoR and leads to the smaller magnitude of $z_{\rm obs}=5.5$ for the latter two scenarios. We show $P_{21}^{\rm patchy}$ as a function of the cosine of the line-of-sight angle $\mu$ in Figure~\ref{fig:p_patchy_mu}. Note that here the $\mu$-dependence comes from the velocity effects rather than any intrinsic anisotropy in the $\Xi$ field.

To further understand the inhomogeneous reionization impact, we plot the ratio of $|P_{21}^{\rm patchy}|$ to the fiducial 21 cm power spectrum $P_{21}^{\rm fid}$ in Figure~\ref{fig:reion_mem}. The ratio peaks at wavenumber $k \sim 0.07 \ \textup{Mpc}^{-1}$. In comparison to the fiducial 21 cm power spectrum, the patchy reionization indeed leave larger imprints on higher redshifts, i.e., the ratio peaks at $z_{\rm obs}=5.5$ for all 3 reionization scenarios and attenuates at lower redshifts. Nonetheless, for delayed reionization the imprints observed at $z_{\rm obs}=5.5$ and $z_{\rm obs}=5.0$ are closer, as the two curves stick in the Late Reionization scenario. 

\subsection{Parameter shifts}

We compute the cosmological parameter shifts due to inhomogeneous reionization for SKA1-LOW and two PUMA observation scenarios. To reduce the thermal noise that is linearly proportional to the survey area, this mode has to choose a small survey area, as shown in Table~\ref{tab:pu-sk} that $f_{\rm sky}=0.0024$. On the other hand, the survey area is also proportional to the number of survey modes so it is positively correlated to the constraining power for cosmological parameters. We find that the limited survey area suppresses SNR of cosmology. Then the cosmological parameter shifts due to inhomogeneous reionization will not be significant with the sensitivity of this observing mode. Regarding the presence of the two competing factors in the choice of survey area for SKA1-LOW, a survey strategy that would like to investigate the inhomogeneous reionization effect may require a fine-tuned survey strategy. We present the forecast $1\sigma$ sensitivity, the parameter shifts, the percentage of shifts with respect to the fiducial parameter value, and the ratio of shifts to uncertainty with all three reionization scenarios in Table~\ref{table:shifts}. The two sub-tables represent \{PUMA, $t_{\rm int}=1000\,\rm hours$\}, and \{PUMA, $t_{\rm int}=5\,\rm years$\} respectively. We show the parameter shifts results assuming the full wedge information could be rendered in Table \ref{table:shifts}. 

The impact of inhomogeneous reionization on 21 cm IM measurement of post-reionization cosmological parameters leads to biases much greater than the statistical uncertainties. For example, for the \{PUMA, $t_{\rm int}$ = 1000 hours\} case with the Planck Reionization scenario (lower panel of Table~\ref{table:shifts}), there are shifts from sub-percent to tens of percent level for most cosmological parameters, while 6 out of 11 parameters we investigate have shifts $>1\sigma$. The smallest fractional shift is 0.9\% for baryon density parameter $\Omega_{\rm b}h^2$, but even then the effect is $>1\sigma$. The sum of the neutrino masses is shifted by $-0.025\,$eV, which is both a large bias relative to the statistical error ($-2.9\sigma$) and comparable to the target sensitivity for distinguishing the neutrino hierarchy.

Comparing the parameter shifts of the two reionization scenarios, it is noticeable that the imprints of inhomogeneous reionization on the post-reionization measurement of some cosmological parameters decay along with time after the end of reionization. Take the results for the Hubble constant $h$ in the first row of Table~\ref{table:shifts} \{PUMA, $t_{\rm int}=$ 5 years\} scenario, for example, with the fiducial value $h=0.6774$, the mean shift is -0.0156 for Early Reionization, -0.0220 for Planck Reionization, -0.0299 for Late Reionization, indicating earlier ended reionization leads to smaller parameter shift. This trend usually applies to other parameters as well in the tables. However, for certain parameters such as $\tau_{\nu}$ and $\sum m_{\nu}$, the shifts do not change monotonically from Early Reionization to Late Reionization. This is likely because the fit is done in a high-dimensional cosmological parameter space, with biases occurring along all of the principal axes of the error ellipse. Some of these biases project into a positive bias in (for example) h, while others project into a negative direction. This can lead to non-monotonic behavior when changing the bias in the parameter vector by a different percentage for different $k, z_{\rm obs}$ bins, or when changing the relative weighting of CMB and 21 cm information (as occurs when we extend PUMA from 1000 hours to 5 years).


As for comparison between different experimental scenarios between tables, the parameter shifts are affected by the thermal noise variations from different instruments and also by the aforementioned high-dimensional parameter space projection phenomenon. Reducing integration time $t_{\rm int}$ for PUMA from 5 years to 1000 hours, the parameter uncertainties increase and the shifts due to inhomogeneous reionization drop dramatically. For the Planck Reionization scenario, the shift of $h$ is -0.0220 for \{PUMA, $t_{\rm int}=5\,\rm years$\} while it is just 0.0096 for \{PUMA, $t_{\rm int}=1000\,\rm hours$\} scenario. This is partly attributed to the thermal noise power spectrum $P_{\rm N}\propto 1/t_{\rm int}$ in Eq.~(\ref{eq:p_thermal}): the effect of inhomogeneous reionization on the total power spectrum is less noticeable when the thermal noise is larger. But the shifts also flip sign between the two scenarios, which is likely due to competing biases in projecting from the principal axes of the error ellipsoid down to a single parameter such as $h$.

Comparing the results when contaminated wavemodes in the ``wedge'' are removed in Table~\ref{table:shifts} and when these modes are taken into account in Appendix~\ref{appendix:wdgoff}, Table~\ref{table:shifts_wdgon}, the impact both in absolute shift and in numbers of $\sigma$ are noticeably greater when more information in the ``wedge'' is available. In the \{PUMA, $t_{\rm int}=1000\,\rm hours$\} scenario, when wedge effect is turned off, $\sigma$ of $h$ reduces from 0.010299 to 0.007638 and that of $\sum m_{\nu}$ reduces from 0.038635 to 0.02208, etc. However, the dilemma is that while the cosmological parameter constraints become tighter with access to information from the ``wedge'', the impact from inhomogeneous reionization becomes more significant as well. Shifts of the ``wedge-off' case in Table~\ref{table:shifts} are generally increased compared to the ``wedge-on'' case in Table~\ref{table:shifts_wdgon} and more parameters shifts go beyond 1$\sigma$.

\section{Discussion}
\label{sec:disc}

This work investigates the impact of inhomogeneous reionization on post-reionization 21 cm IM measurement of cosmological parameters $\{h,\,\Omega_{\rm b}h^2$,\,$\Omega_{\rm c}h^2$,\,$\tau$,\, $\sum m_{\nu}$,\,$n_{\rm s}, \,10^9A_{\rm s},\,b_{\rm HI},\,10^3\Omega_{\rm HI}\}$. The passage of ionization fronts changes the temperature of the IGM instantly and leaves imprints on matter distribution as well by heating gas to escape from shallow gravitational wells of mini-halos, thus reducing the abundance of small-scale baryon structure. With discrete ionizing sources, reionization happens inhomogeneously through the Universe. As the relaxation time of IGM to react to the passage of ionization fronts is comparative to the duration of EoR, the inhomogeneous history of reionization gives rise to inhomogeneous post-reionization states of IGM, which could bias the measurement of $\Lambda$CDM model parameters. In this work, we incorporate small-scale baryon structure response to ionization fronts into the patchy reionization history featured in large scales 
in the calculation of post-reionization \HI\ distribution by the hybrid method described in \S\ref{sec:sims}. To understand the imprints of patchy reionization on post-reionization power spectrum, we show $P_{21}^{\rm patchy}$ with various $z_{\rm obs}$ and three reionization scenarios in Figure~\ref{fig:p_patchy} and the ratio of its amplitude to the total 21 cm power spectrum $|P_{21}^{\rm patchy}|/P_{21}^{\rm fid}$ in Figure~\ref{fig:reion_mem}. For Planck Reionization scenario, at the peak the reionization memory makes contribution $\sim 25\%$ at $z=5.5$ and $\sim 10\%$ at $z=3.5$ to the total power spectrum. We find that the impact of patchy reionization usually attenuates through time after the end of reionization. But the violent relaxation right after reionization adds a complication to the post-reionization state of IGM, and $P_{21}^{\rm patchy}$ in some cases does not peak at higher observation redshift $z_{\rm obs}$, as shown in the middle and right panel of Figures~\ref{fig:p_patchy} and \ref{fig:p_patchy_mu}.

\begin{table*}
\caption{Summary of cosmological parameter fiducial value, forecast $1\sigma$ sensitivity, and parameter shifts due to inhomogeneous reionization with three reionization scenarios assuming that modes in the ``wedge'' can be recovered. We show the 1000 times parameter shifts in the columns of $\rm Shifts \times10^3$, the percent of shifts to the fiducial value in the column of Shift \%, and the ratio of shifts to the forecast $1\sigma$ in the column of Shift/$\sigma$. The two experimental scenarios \{PUMA, $t_{\rm int}=$ 1000 hours\} and \{PUMA, $t_{\rm int}=$ 5 years\}  are displayed here.} 

\center{\large PUMA, $t_{\rm int}=1000$ hours} 
\begin{tabular}{cccccccccccc}
\hline
\hline
Parameter & Fiducial & Forecast 1$\sigma$  & \multicolumn3c{Early} & \multicolumn3c{Planck} & \multicolumn3c{Late}\\
~ & ~ & ~ &  Shift $\times 10^3$  & \!\!Shift \%\!\! & \!Shift/$\sigma$\! & Shift $\times 10^3$  & \!\!Shift \%\!\! & \!Shift/$\sigma$\! & Shift $\times 10^3$ & \!\!Shift \%\!\! & \!Shift/$\sigma$\! \\
\cmidrule(lr){4-6}\cmidrule(lr){7-9}\cmidrule(lr){10-12} 
\hline\\
h & 0.6774 & 0.007638  & $6.7 \pm 1.4$ & 1.0 & 0.9 & $9.6 \pm 1.8$ & 1.4 & 1.3 & $12.6 \pm 2.0$ & 1.9 & 1.7
\\
$\Omega_b h^2$ & 0.0223 & 0.000235  & $0.13 \pm 0.07$ & 0.6 & 0.6 & $0.19 \pm 0.08$ & 0.9 & 0.8 & $0.26 \pm 0.09$ & 1.2 & 1.1
\\
$\Omega_c h^2$ & 0.1188 & 0.001489  & $-1.0 \pm 0.2$ & -0.8 & -0.7 & $-1.6 \pm 0.3$ & -1.3 & -1.1 & $-2.4 \pm 0.4$ & -2.0 & -1.6
\\
$\sum m_{\nu}$ & 0.1940 & 0.022122  & $-17.8 \pm 3.4$ & -9.2 & -0.8 & $-19.2 \pm 4.7$ & -9.9 & -0.9 & $-15.1 \pm 6.1$ & -7.8 & -0.7
\\
$A_s\times10^9$ & 2.1420 & 0.047047  & $29.2 \pm 3.5$ & 1.4 & 0.6 & $34.4 \pm 4.6$ & 1.6 & 0.7 & $32.9 \pm 6.0$ & 1.5 & 0.7
\\
$n_s$ & 0.9667 & 0.005909  & $5.7 \pm 0.8$ & 0.6 & 1.0 & $6.6 \pm 1.0$ & 0.7 & 1.1 & $5.8 \pm 1.2$ & 0.6 & 1.0
\\
$\tau_{\rm re}$ & 0.0660 & 0.010809  & $8.3 \pm 0.9$ & 12.6 & 0.8 & $10.0 \pm 1.2$ & 15.1 & 0.9 & $10.0 \pm 1.6$ & 15.1 & 0.9
\\
$b_{\rm HI,1}$ & 2.8200 & 0.034625  & $-76.1 \pm 1.1$ & -2.7 & -2.2 & $-109.1 \pm 1.3$ & -3.9 & -3.2 & $-143.7 \pm 1.6$ & -5.1 & -4.2
\\
$b_{\rm HI,2}$ & 3.1800 & 0.045894  & $0.7 \pm 12.6$ & 0.0 & 0.0 & $-40.4 \pm 15.9$ & -1.3 & -0.9 & $-127.0 \pm 19.8$ & -4.0 & -2.8
\\
$\Omega_{\rm HI,1}\times10^3$ & 1.1800 & 0.036446  & $-53.7 \pm 4.9$ & -4.5 & -1.5 & $-67.6 \pm 6.4$ & -5.7 & -1.9 & $-72.9 \pm 7.7$ & -6.2 & -2.0
\\
$\Omega_{\rm HI,2}\times10^3$ & 0.9800 & 0.139335  & $-62.3 \pm 4.3$ & -6.4 & -0.4 & $-70.2 \pm 5.5$ & -7.2 & -0.5 & $-61.2 \pm 7.1$ & -6.2 & -0.4
\\
\hline\hline
\end{tabular}

\center{\large PUMA, $t_{\rm int}=5$ years} 
\begin{tabular}{cccccccccccc}
\hline
\hline
Parameter & Fiducial & Forecast 1$\sigma$  & \multicolumn3c{Early} & \multicolumn3c{Planck} & \multicolumn3c{Late}\\
~ & ~ & ~ &  Shift $\times 10^3$  & \!\!Shift \%\!\! & \!Shift/$\sigma$\! & Shift $\times 10^3$  & \!\!Shift \%\!\! & \!Shift/$\sigma$\! & Shift $\times 10^3$ & \!\!Shift \%\!\! & \!Shift/$\sigma$\! \\
\cmidrule(lr){4-6}\cmidrule(lr){7-9}\cmidrule(lr){10-12} 
\hline\\
h & 0.6774 & 0.002208  & $-15.6 \pm 0.4$ & -2.3 & -7.1 & $-22.0 \pm 0.6$ & -3.2 & -9.9 & $-29.9 \pm 0.7$ & -4.4 & -13.5
\\
$\Omega_b h^2$ & 0.0223 & 0.000074  & $-0.37 \pm 0.05$ & -1.6 & -5.0 & $-0.48 \pm 0.07$ & -2.2 & -6.6 & $-0.61 \pm 0.07$ & -2.7 & -8.3
\\
$\Omega_c h^2$ & 0.1188 & 0.000376  & $2.7 \pm 0.0$ & 2.2 & 7.1 & $3.5 \pm 0.1$ & 3.0 & 9.4 & $4.4 \pm 0.1$ & 3.7 & 11.7
\\
$\sum m_{\nu}$ & 0.1940 & 0.003115  & $-14.2 \pm 1.5$ & -7.3 & -4.6 & $-14.7 \pm 2.0$ & -7.6 & -4.7 & $-10.8 \pm 2.4$ & -5.6 & -3.5
\\
$A_s\times10^9$ & 2.1420 & 0.025592  & $21.7 \pm 5.6$ & 1.0 & 0.8 & $28.1 \pm 7.7$ & 1.3 & 1.1 & $36.4 \pm 9.7$ & 1.7 & 1.4
\\
$n_s$ & 0.9667 & 0.001344  & $20.6 \pm 7.5$ & 2.1 & 15.4 & $14.7 \pm 10.2$ & 1.5 & 11.0 & $-7.4 \pm 13.2$ & -0.8 & -5.5
\\
$\tau_{\rm re}$ & 0.0660 & 0.003505  & $7.4 \pm 1.1$ & 11.1 & 2.1 & $8.3 \pm 1.5$ & 12.6 & 2.4 & $8.2 \pm 1.9$ & 12.4 & 2.3
\\
$b_{\rm HI,1}$ & 2.8200 & 0.003094  & $-66.1 \pm 0.6$ & -2.3 & -21.4 & $-94.3 \pm 0.8$ & -3.3 & -30.5 & $-124.5 \pm 1.1$ & -4.4 & -40.2
\\
$b_{\rm HI,2}$ & 3.1800 & 0.008609  & $-109.4 \pm 2.9$ & -3.4 & -12.7 & $-161.2 \pm 3.9$ & -5.1 & -18.7 & $-224.1 \pm 4.9$ & -7.0 & -26.0
\\
$\Omega_{\rm HI,1}\times10^3$ & 1.1800 & 0.010137  & $42.0 \pm 2.8$ & 3.6 & 4.1 & $64.2 \pm 3.8$ & 5.4 & 6.3 & $96.7 \pm 4.9$ & 8.2 & 9.5
\\
$\Omega_{\rm HI,2}\times10^3$ & 0.9800 & 0.007329  & $-54.7 \pm 2.4$ & -5.6 & -7.5 & $-65.4 \pm 3.1$ & -6.7 & -8.9 & $-65.2 \pm 3.7$ & -6.7 & -8.9
\\
\hline\hline
\end{tabular}

\label{table:shifts}
\end{table*}

We present the cosmological parameter shifts results in Table. \ref{table:shifts} for observation scenarios \{PUMA, $t_{\rm int}=1000\,\rm hours$\} and \{PUMA, $t_{\rm int}=5\,\rm years$\} respectively. For \{PUMA, $t_{\rm int}=$ 5 years, Planck Reionization\} scenario, the shifts of parameters range from sub-percent (-0.1\% for $n_{\rm s}$) to tens of percent of their fiducial values, while most of them are order of magnitude larger than the detection sensitivity. These considerable parameter shifts show that the inhomogeneous reionization is a considerable issue as a systematic effect on post-reionization 21 cm IM measurement of $\Lambda$CDM model. For the other observation scenario \{PUMA, $t_{\rm int}=1000\,\rm hours$\}, the inhomogeneous reionization effect induced shifts are smaller. But this is because the thermal noise dominates the systematics due to patchy reionization rather than the latter observation strategy is more optimal.

The results of this work show that the inhomogeneous reionization should be a concern of post-reionization 21 cm probe of cosmological parameters. For future 21 cm IM experiments aiming at probing post-reionization high-redshift Universe like PUMA, it's worth taking this effect into account and slightly push the detection to higher redshift to help calibrate and remove this effect. A scratch of ongoing reionization process in 21 cm observation would improve our knowledge of the patchy reionization process and more comprehensive understanding of evolution of IGM state. As for theoretical mitigation strategy, a possible way towards resolving this issue is by parametrically taking the patchy reionization into account and marginalizing over its parameters when extracting cosmological parameters from observational data \citep{2023MNRAS.520.4853M}. On the other side, we emphasize that originated from the discreteness and characteristics of ionizing sources, the patchy reionization effect itself could be regarded as the information to study the astrophysics that governs reionization \citep{2021MNRAS.508.1262M}. Also, as in Eq.~(\ref{eq:P_21}), the fiducial 21 cm power spectrum and the patchy reionization power spectrum term have different angular dependence originated from redshift distortion. The ratio of $P_{21}^{\rm patchy}$ to $P_{21}^{\rm fid}$ shown in Figure. \ref{fig:reion_mem} thus in principle varies with respect to $\mu$, which could be used in future work to inversely calibrate patchy reionization effect. We don't include any parameter regarding the astrophysics of EoR in our forecast framework. There should be gains to investigate some important astrophysical parameters but considerable amounts of new simulations are needed to achieve this.

There are several simplifications and approximations included in our calculation framework. First, in principle the bias parameter $b_{\rm HI}$ and \HI\ abundance parameter $\Omega_{\rm HI}$ should vary with respect to different cosmology sets. Under the framework of Fisher matrix, to calculate the shift of a specific parameter we marginalize over other parameters. Nonetheless, it is possible to have additional information like priors of the bias $b_{\rm HI}$ and $\Omega_{\rm HI}$ if their relation to cosmology background could be better modeled. Second, our semi-analytical approach of incorporating small-scale structure reaction into large-scale inhomogeneous reionization in this work could lose some correlation between the scales of the {\sc 21cmFAST} box and the small GADGET-2 box. It would require a large box simulation of reionization with subgrid modeling analogous to \citealt{2006MNRAS.366..689C} to capture the correlation in the range of scales we may miss in this work. 
Last but not least, some recent work \citep{2015MNRAS.447.3402B, 2019MNRAS.485L..24K,2020MNRAS.491.1736K} attribute the deep absorption trough observed in the \lya\ forest down to the redshift 5.5 to reionization completes as late as $z=5.3$. We do not include the ultra delayed reionization scenario in our investigation of inhomogeneous reionization result. However, as shown in Figures \ref{fig:p_patchy} and \ref{fig:reion_mem}, the impact of patchy reionization on post-reionization 21 cm signal is sensitive to the time reionization ends, future work should study the delayed reionization case. 

In light of the importance of thoroughly understanding EoR to related cosmology and astrophysics studies shown in this work and previous studies, there have been many observational endeavors towards a better understanding of the EoR spans from large cosmological surveys to point source searches at high-redshifts \citep{fan_observational_2006,2016ARA&A..54..313M}. The classic work of high-$z$ \lya\ forest spectra observation indicate the end of reionization at $z\approx6$ with dramatic increase of IGM opacity \citep{2002AJ....123.1247F, 2006AJ....132..117F}. But some recent work \citep{2015MNRAS.447.3402B, 2019MNRAS.485L..24K,2020MNRAS.491.1736K} with observation of deep absorption trough down to redshift 5.5 in high-$z$ quasar spectra suggest reionization could end as late as $z\approx5.3$. The cosmic microwave background (CMB) constrains the global reionization history by large-scale polarization anisotropies from the average Thomson scattering optical depth through reionization, \textit{Planck} 2018 ``$TT+TE+EE+{\rm low}E+{\rm lensing}$'' gives a reionization midpoint at $z_{\rm mid}=7.74$ \citep{2020A&A...641A...6P}. Secondary anisotropy of CMB is expected to be used to constrain patchy reionization, as it's sourced by the kinetic Sunyaev-Zel'dovich (kSZ) \citep{1980MNRAS.190..413S} effect due to photons Doppler scattering off the relative motions of ionized structures \citep{2017MNRAS.465.4838G,2020A&A...640A..90G,2022arXiv220304337C}. The damping wing at the redside of \lya\ spectra due to absorption of neutral IGM is also a probe of state of ionization of IGM \citep{2018ApJ...864..142D} and can be used to constrain inhomogeneous reionization \citep{mesinger_ly_2008}. The successfully launched James Webb Space Telescope (JWST) \citep{2006SSRv..123..485G} is expected to shed light on the first luminous sources that leads to reionization and enhance our knowledge of reionization history. The 21 cm IM is capable of mapping the reionization of IGM and thus has the potential to yield vastly more information of reionization than any other probe. Though it is a constant challenge to dig up the buried \HI\ 21 cm signal from its several order of magnitude larger foreground noise, 21 cm can detect the high-redshift Universe with large survey volume compared to galaxy or \lya\ surveys, the tracer densities of which is largely reduced at higher redshifts. Furthermore, along with other novel line-intensity mapping probes tracing the star-forming galaxies that create the ionizing photons such as CO and C{\,\sc ii} lines \citep{2017arXiv170909066K,2020ApJ...892...51C, 2021ApJ...909...51Z, 2021ApJ...915...33S}, \HI\ 21 cm IM is promising on unraveling abundant cosmological and astrophysical information about the EoR. 
The results of this paper imply that these lines of work are important not only to the astrophysical understanding of reionization, but also to probing fundamental physics using the post-reionization hydrogen intensity mapping signal.

\section*{Acknowledgements}
We are grateful to Yi Mao, Yao Zhang and Chun-Hao To for helpful suggestions and comments. We thank the anonymous referee for comments regarding the 21 cm experiments configuration and ``wedge'' issue. CMG is thankful to Alpha-Cen and the Cenca Bridge undergraduate remote internship program for their efforts on connecting undergraduate students from Central America and the Caribbean with researchers in other regions. HL and CMH were supported by OSU Presidential Fellowship, NASA award 15-WFIRST15-0008, Simons Foundation award 60052667, and the David \& Lucile Packard Foundation award 2021-72096.  PMC was supported by NSFC grant No.~12050410236, the Major Key Project of PCL, and the Tsinghua Shui Mu Scholarship. We acknowledge the Pitzer Cluster at the Ohio Supercomputing Center \citep{OhioSupercomputerCenter1987} and Tsinghua Astrophysics High-Performance Computing platform at Tsinghua University for providing computational and data storage resources that have contributed to the research results reported within this paper.

\section*{Data Availability}
The data underlying this article are available in a GitHub repository, appropriate link is given in the manuscript.



\bibliographystyle{mnras}
\bibliography{main} 

\newpage
\appendix

\section{Cosmological parameters shifts with ``wedge'' modes removed}
\label{appendix:wdgoff}

In this appendix, we show the cosmological parameter shifts due to inhomogeneous reionization (analogous to Table~\ref{table:shifts}) but assuming that the contaminated ``wedge'' modes can not be recovered for PUMA. We find that the shifts are smaller in absolute magnitude and relative to the forecast statistical uncertainty (shift/$\sigma$).

\begin{table*}
\caption{Summary of cosmological parameter fiducial value, forecast $1\sigma$ sensitivity, and parameter shifts due to inhomogeneous reionization with three reionization scenarios when modes in the ``wedge'' are excluded. We show the 1000 times parameter shifts in the columns of $\rm Shifts \times10^3$, the percent of shifts to the fiducial value in the column of Shift \%, and the ratio of shifts to the forecast $1\sigma$ in the column of Shift/$\sigma$. The two experimental scenario \{PUMA, $t_{\rm int}=$ 1000 hours\} and \{PUMA, $t_{\rm int}=$ 5 years\}  are displayed here.}

\center{\large PUMA, $t_{\rm int}=1000$ hours} 
\begin{tabular}{cccccccccccc}
\hline
\hline
Parameter & Fiducial & Forecast 1$\sigma$  & \multicolumn3c{Early} & \multicolumn3c{Planck} & \multicolumn3c{Late}\\
~ & ~ & ~ &  Shift $\times 10^3$  & \!\!Shift \%\!\! & \!Shift/$\sigma$\! & Shift $\times 10^3$  & \!\!Shift \%\!\! & \!Shift/$\sigma$\! & Shift $\times 10^3$ & \!\!Shift \%\!\! & \!Shift/$\sigma$\! \\
\cmidrule(lr){4-6}\cmidrule(lr){7-9}\cmidrule(lr){10-12} 
\hline
h & 0.6774 & 0.010299  & $3.5 \pm 1.4$ & 0.5 & 0.3 & $4.9 \pm 1.8$ & 0.7 & 0.5 & $6.5 \pm 1.9$ & 1.0 & 0.6
\\
$\Omega_b h^2$ & 0.0223 & 0.000268  & $0.06 \pm 0.04$ & 0.2 & 0.2 & $0.09 \pm 0.05$ & 0.4 & 0.3 & $0.13 \pm 0.06$ & 0.6 & 0.5
\\
$\Omega_c h^2$ & 0.1188 & 0.001908  & $-0.5 \pm 0.3$ & -0.4 & -0.2 & $-0.8 \pm 0.4$ & -0.7 & -0.4 & $-1.3 \pm 0.5$ & -1.1 & -0.7
\\
$\sum m_{\nu}$ & 0.1940 & 0.038635  & $-12.4 \pm 3.1$ & -6.4 & -0.3 & $-11.9 \pm 4.1$ & -6.1 & -0.3 & $-6.1 \pm 5.0$ & -3.1 & -0.2
\\
$A_s\times10^9$ & 2.1420 & 0.054381  & $12.1 \pm 2.4$ & 0.6 & 0.2 & $15.0 \pm 3.1$ & 0.7 & 0.3 & $15.8 \pm 3.6$ & 0.7 & 0.3
\\
$n_s$ & 0.9667 & 0.007255  & $2.4 \pm 0.5$ & 0.2 & 0.3 & $2.9 \pm 0.7$ & 0.3 & 0.4 & $2.9 \pm 0.8$ & 0.3 & 0.4
\\
$\tau_{\rm re}$ & 0.0660 & 0.013115  & $3.5 \pm 0.8$ & 5.4 & 0.3 & $4.5 \pm 1.0$ & 6.8 & 0.3 & $4.9 \pm 1.1$ & 7.4 & 0.4
\\
$b_{\rm HI,1}$ & 2.8200 & 0.385024  & $133.4 \pm 2.3$ & 4.7 & 0.3 & $176.2 \pm 3.0$ & 6.2 & 0.5 & $213.5 \pm 4.1$ & 7.6 & 0.6
\\
$b_{\rm HI,2}$ & 3.1800 & 0.475164  & $45.0 \pm 14.7$ & 1.4 & 0.1 & $-12.9 \pm 18.7$ & -0.4 & -0.0 & $-101.3 \pm 20.1$ & -3.2 & -0.2
\\
$\Omega_{\rm HI,1}\times10^3$ & 1.1800 & 0.131434  & $-104.2 \pm 4.3$ & -8.8 & -0.8 & $-136.6 \pm 5.6$ & -11.6 & -1.0 & $-160.0 \pm 6.9$ & -13.6 & -1.2
\\
$\Omega_{\rm HI,2}\times10^3$ & 0.9800 & 1.981524  & $-68.4 \pm 3.2$ & -7.0 & -0.0 & $-71.5 \pm 4.0$ & -7.3 & -0.0 & $-62.2 \pm 4.2$ & -6.3 & -0.0
\\
\hline\hline
\end{tabular}

\center{\large PUMA, $t_{\rm int}=5$ years} 
\begin{tabular}{cccccccccccc}
\hline
\hline
Parameter & Fiducial & Forecast 1$\sigma$  & \multicolumn3c{Early} & \multicolumn3c{Planck} & \multicolumn3c{Late}\\
~ & ~ & ~ &  Shift $\times 10^3$  & \!\!Shift \%\!\! & \!Shift/$\sigma$\! & Shift $\times 10^3$  & \!\!Shift \%\!\! & \!Shift/$\sigma$\! & Shift $\times 10^3$ & \!\!Shift \%\!\! & \!Shift/$\sigma$\! \\
\cmidrule(lr){4-6}\cmidrule(lr){7-9}\cmidrule(lr){10-12} 
\hline
h & 0.6774 & 0.004255  & $-13.4 \pm 2.2$ & -2.0 & -3.1 & $-17.5 \pm 3.0$ & -2.6 & -4.1 & $-21.0 \pm 3.5$ & -3.1 & -4.9
\\
$\Omega_b h^2$ & 0.0223 & 0.000157  & $-0.33 \pm 0.10$ & -1.5 & -2.1 & $-0.43 \pm 0.13$ & -1.9 & -2.7 & $-0.51 \pm 0.15$ & -2.3 & -3.3
\\
$\Omega_c h^2$ & 0.1188 & 0.000784  & $3.3 \pm 0.4$ & 2.8 & 4.2 & $4.1 \pm 0.5$ & 3.5 & 5.2 & $4.6 \pm 0.6$ & 3.9 & 5.8
\\
$\sum m_{\nu}$ & 0.1940 & 0.007700  & $-24.1 \pm 1.5$ & -12.4 & -3.1 & $-30.1 \pm 2.0$ & -15.5 & -3.9 & $-33.8 \pm 2.5$ & -17.4 & -4.4
\\
$A_s\times10^9$ & 2.1420 & 0.048719  & $-13.7 \pm 3.9$ & -0.6 & -0.3 & $-26.1 \pm 5.3$ & -1.2 & -0.5 & $-43.2 \pm 6.7$ & -2.0 & -0.9
\\
$n_s$ & 0.9667 & 0.002834  & $8.1 \pm 2.9$ & 0.8 & 2.9 & $3.3 \pm 3.8$ & 0.3 & 1.2 & $-10.7 \pm 5.0$ & -1.1 & -3.8
\\
$\tau_{\rm re}$ & 0.0660 & 0.011015  & $-4.0 \pm 1.0$ & -6.0 & -0.4 & $-7.8 \pm 1.4$ & -11.8 & -0.7 & $-13.5 \pm 1.8$ & -20.4 & -1.2
\\
$b_{\rm HI,1}$ & 2.8200 & 0.049682  & $119.3 \pm 1.0$ & 4.2 & 2.4 & $164.2 \pm 1.4$ & 5.8 & 3.3 & $215.0 \pm 1.9$ & 7.6 & 4.3
\\
$b_{\rm HI,2}$ & 3.1800 & 0.039029  & $69.2 \pm 0.6$ & 2.2 & 1.8 & $33.3 \pm 1.2$ & 1.0 & 0.9 & $-37.2 \pm 2.0$ & -1.2 & -1.0
\\
$\Omega_{\rm HI,1}\times10^3$ & 1.1800 & 0.026092  & $-7.2 \pm 5.9$ & -0.6 & -0.3 & $-12.0 \pm 7.8$ & -1.0 & -0.5 & $-19.1 \pm 8.9$ & -1.6 & -0.7
\\
$\Omega_{\rm HI,2}\times10^3$ & 0.9800 & 0.147005  & $-98.0 \pm 1.9$ & -10.0 & -0.7 & $-111.9 \pm 2.4$ & -11.4 & -0.8 & $-108.9 \pm 2.9$ & -11.1 & -0.7
\\
\\
\hline\hline
\end{tabular}
\label{table:shifts_wdgon}
\end{table*}

\end{document}